\documentclass[amsmath,amssymb,twocolumn,pre,floatfix]{revtex4}

\def\E{{\rm e}}
\def\I{{\rm i}}
\usepackage{color}
\usepackage{graphicx}
\usepackage{dcolumn}
\usepackage{times}
\usepackage{soul}

\usepackage{afterpage}

\begin{document}

\title{Programmable Filaments and Textiles}
\author{A.~P. Zakharov and L.~M. Pismen}
\affiliation{Technion -- Israel Institute of Technology, Haifa 32000, Israel} 

\begin{abstract}
We analyze the various morphing structures obtained by actuating Janus filaments comprising driven and passive sectors and textiles incorporating driven and passive filaments. Transitions between alternative shapes and coexistence of absolutely stable and metastable states within a certain range of relative extension upon actuation are detected both in Janus rings and textiles. Both single filaments and textiles can be reverse designed to bend into desired shapes by controlling both the size and orientation of driven sectors.
\end{abstract}
 
 \maketitle
 
\section{Introduction} 

A major challenge in the development of soft robotics \cite{robot} and smart textiles \cite{smart} is programming of desired shapes attainable upon actuation. Common programmable soft materials reshaping under the influence of the various external factors include hydrogels swelling or shrinking by imbibing or expelling the solvent  \cite{Balazs} and liquid crystal elastomers (LCE), made of cross-linked polymeric chains with embedded mesogenic structures, which reshape owing to changes of the nematic order parameter and the director orientation that can be pre-programmed and controlled by the various physical and chemical agents, such as heat, light, electric or magnetic field  \cite{Warner}. Most studies of actuation of LCE explored deformation of thin pre-patterned sheets \cite{sharon,mostajeran}. The inverse problem is patterning of a flat sheet that would buckle into a desired shape upon actuation \cite{Aharoni}.

\begin{figure}[b]
\begin{tabular}{cc}
		(a)&(b)\\ 
		\includegraphics[width=.21\textwidth]{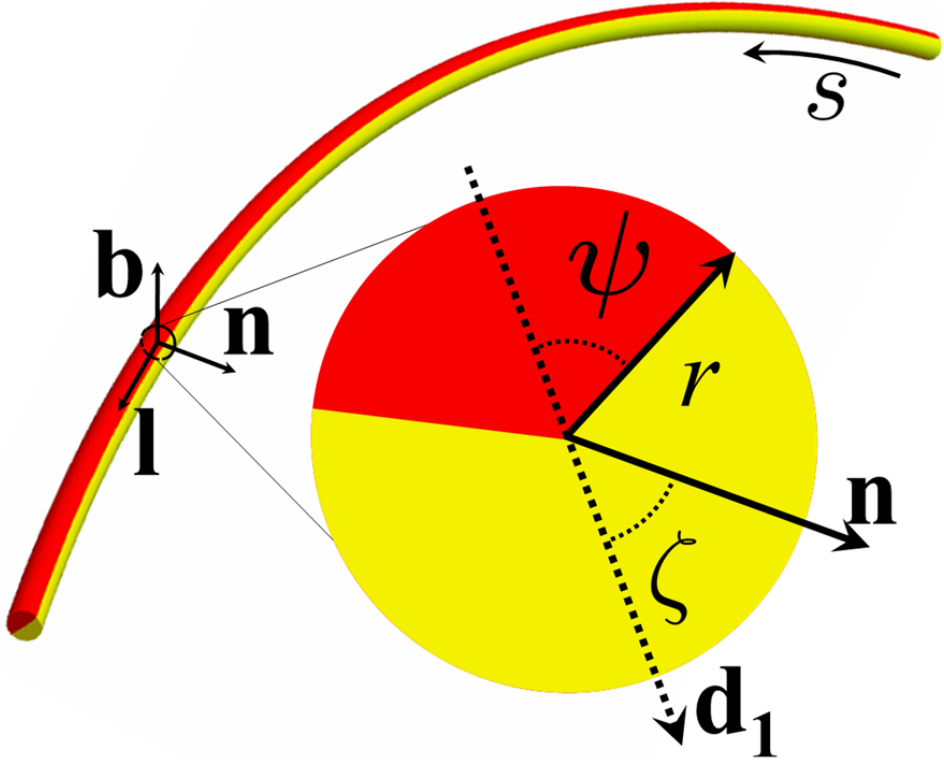}
		&\includegraphics[width=.23\textwidth]{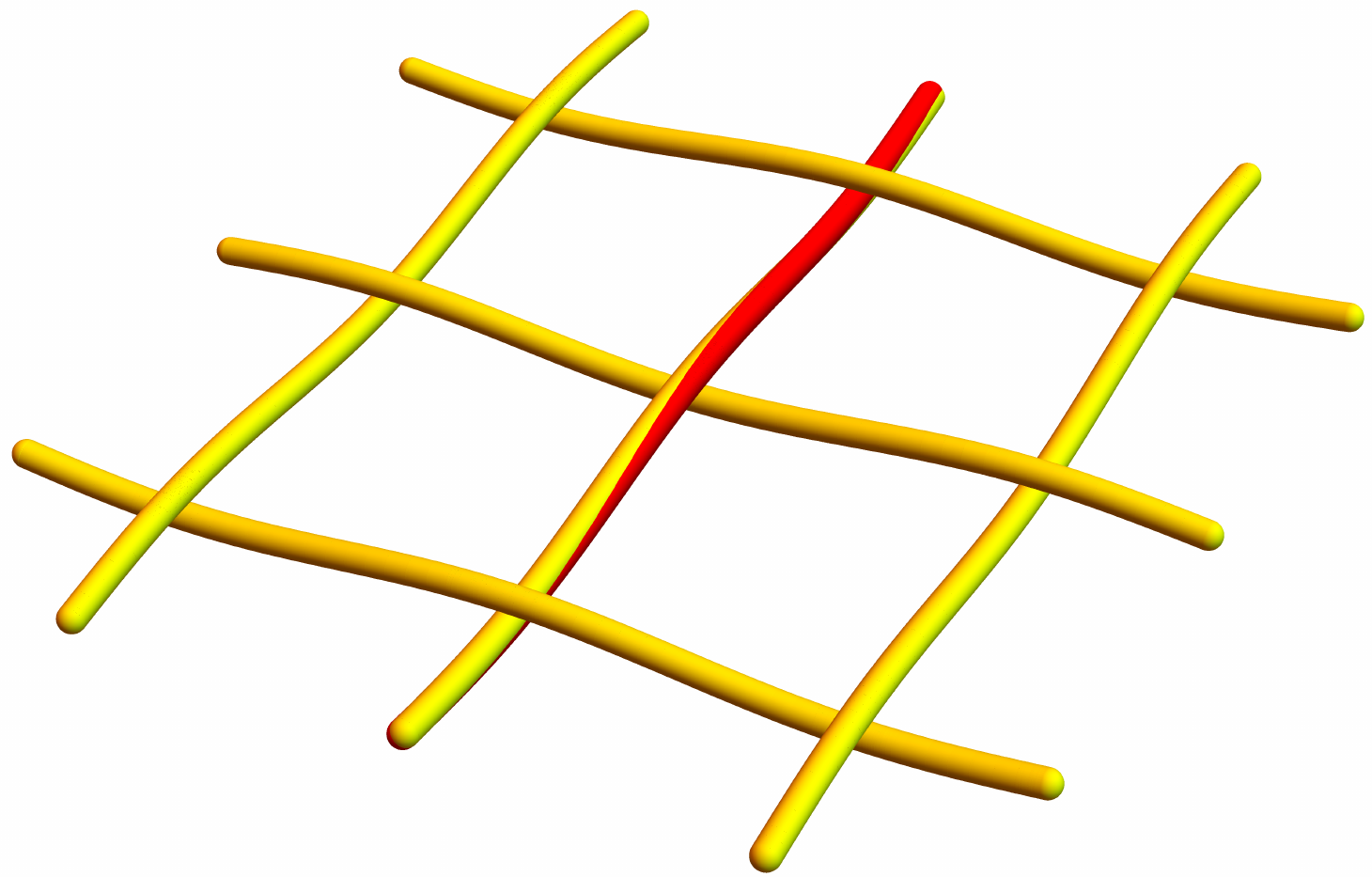}
\end{tabular}
\caption{(a) The cross-section geometry of a Janus filament with a mismatched orientation of the Frenet normal vector \textbf{n} and the material basis. (b) A piece of textile embroidered by a Janus filament.}  
\label{sketch} 
\end{figure}

Bending and twisting of intrinsically curved filaments has been largely studied in application to ``birods", composite filaments glued of prestressed and unstressed stripes, as reviewed by Goriely \cite{Goriely06}. Similar forms with  intrinsic curvature and twist exist in Nature where they emerge during plant growth \cite{GorielyTabor,GorielyGoldstein}. Intrinsic curvature can be generated in composite filaments  -- \emph{Janus filaments} (Fig.~\ref{sketch}a), combining driven (a hydrogel or a longitudinally polarised LCE) strands with a passive component. Such filaments can be fabricated either by 3D-printing or by using two connected extruders with simultaneous melt spinning to generate a filament containing two different materials \cite{Ionov}. Their substantial distinction from prestressed birods stems from the material nature of a cross-sectional inhomogeneity. As a result, such a filament remains under compression even when optimally bent \cite{pre18}, but the mathematical treatment of the bending and twisting is similar in both cases.

Textiles incorporating Janus filaments comprising driven and passive sectors \cite{textile}  (Fig.~\ref{sketch}b) provide far more freedom of creating a variety of shapes than single filaments limited by their single dimension or sheets constrained by the continuity requirements that often lead to creation of sharp edges at locations of topological defects of the order parameter field or at sharp fronts between domains with different properties that may readily arise in the course of actuation \cite{pre17}. Moreover, gliding and bending freedom at nodes of a woven fabric allow for a greater flexibility compared with 3D-printed grids \cite{Maha}. 

In this communication, we carry out a comprehensive investigation of shape changing structures incorporating Janus filaments with the help of both analytical and numerical tools. We start in Sect.~\ref{S2} with defining bending and twisting energy of a single Janus filament. This is followed in Sect.~\ref{S3} by demonstrating the various forms generated by actuating single open-ended Janus filaments, including both analytical and numerical reverse design of desired shapes. In Sect.~\ref{S4}, we study the effect of constraints imposed by closing Janus filaments into rings with sub-optimal curvature. We further study textiles combining Janus and passive filaments, starting with a basic elementary frame structure in Sect.~\ref{S5} and continuing in Sect.~\ref{S6} with morphing a piece of textile embroidered by Janus filaments. In this system, we encounter transitions between alternative shapes and coexistence of absolutely stable and metastable states within a certain range of relative extension upon actuation. Additional constraints and a reacher bifurcation structure are observed in structures incorporating Janus rings Sect.~\ref{S71}. A short Sect.~\ref{S7} gives an idea of reverse numerical design of textiles composed of Janus filaments with variable sizes and orientations of the active sector leading to a variety of desired forms. 

\section{Geometry and energy of Janus filaments} \label{S2}
\subsection{Bending energy} \label{S21}

The elastic energy of a thin filament, whether passive or \emph{driven}, i.e. capable to change its geometry in response to external inputs, is a combination of stretching, bending and twisting energies. Since stretching rigidity is proportional to the cross-sectional area, while the other two rigidities are proportional to  the cross-sectional area squared, a thin filament can be assumed inextensible, and its equilibrium shape is determined by minimizing the sum of bending and twisting energies:
\begin{equation}
\mathcal{F} =  \frac12 A E \int \left( I + J \right) ds, \label{Fe1}
\end{equation}
where $A$ is the cross-sectional area, $E$ is the Young modulus, and $I, \, J$ is the bending and twist momenta, respectively. 

We are particularly interested in \emph{Janus} filaments, both free-standing and elements of woven textiles, combining passive and driven components. We further restrict to filaments with a circular cross-section and a driven component occupying a sector with an angle $2\psi$ (see Fig.~\ref{sketch}a), with the passive component filling the remaining sector. Clearly, the material inhomogeneity breaks the symmetry of the cross-section. One can define therefore at each point on the centerline of the filament, alongside the Frenet frame spanned by the normal \textbf{n}, binormal \textbf{b}, and tangent \textbf{l} vectors, the \emph{material basis} \{${\bf d},\, {\bf d}',\, {\bf l}$\}, where the vector ${\bf d}$ is directed along the midline of the driven sector, and ${\bf d}' ={\bf \l} \times{\bf d}$. Upon actuation, the driven component elongates locally by a factor $ 1+\epsilon, \; |\epsilon| \ll 1$, causing the filament to develop curvature $\kappa=1/R$. Further on, we assume $\epsilon> 0$; the opposite effect of a driven component contracting upon actuation is treated in the same way, differing only by reversing the bending direction.  Our goal is to obtain equilibrium shapes in the absence of external forces, so that, as a simplifying assumption, we neglect inertial as well as gravitational contributions to the overall energy. 

The bending momentum, determined by the total elastic energy per unit cross-sectional area of a thin filament with a radius $r \ll R$ is computed by adding the contributions of the driven ($I_a$) and passive ($I_p$) sectors. In the absence of external constraints, the curvature is directed along the vector ${\bf d}$ in the material frame, which therefore coincides with the Frenet normal vector  \textbf{n}.  Then straightforward computation yields
\begin{align}
I_p &=\frac{1}{\pi r^2}\int_0^r \frac {\rho^3}{R^2} \,d \rho  
\int_{\psi-\pi}^{\pi-\psi}\cos^2 \phi \,d \phi  \notag \\
&=  \frac{(\kappa r)^2}{4\pi}\left(\pi-\psi- \sin \psi \cos \psi\right), \label{Iout} \\
I_a &=\frac{1}{\pi r^2}\int_0^r \rho \,d \rho 
\int_{\pi-\psi}^{\pi+\psi} \left(\frac \rho R \cos \phi -\epsilon\right)^2 d \phi  \label{Iin} \\
 &= \frac{(\kappa r)^2}{4\pi} \left(\psi +\frac 12 \sin 2\psi\right)- \frac{4\kappa r\epsilon}{3\pi}\sin \psi +  \frac{\epsilon^2\psi}{\pi} ,\notag \\
I (\kappa,\psi) &= I_a+I_p = \frac 14  \left(\kappa r\right)^2 
 -\frac {4\,\epsilon}{3 \pi} \kappa  r \sin\psi+\frac {\epsilon^2\psi}{\pi} .
 \label{Ipi}
\end{align}
The equilibrium curvature $\widehat{\kappa}$ is defined by the condition $dI/d\kappa=0$:  
 \begin{equation}
\widehat{\kappa}= \frac 83 \frac{\epsilon}{\pi r}\sin\psi.
\label{khat}
\end{equation}
Clearly, extending by a small fraction $\epsilon \ll 1$ is sufficient to attain a curvature radius $R \gg r$ compatible with the common thin filament approximation. Under these conditions, slight variations of the cross-sectional area can be neglected, and it is inessential whether the driven component is a hydrogel or a semicrystalline polymer enlarging the filament diameter in the same proportion as its length or a LCE proportionally shrinking the cross-section. The common thin filament approximation is applicable at $\kappa r \ll 1$, and therefore we further restrict to small extensions.

Due to interactions among filaments, constraints or external forces, the direction of ${\bf d}$ in material basis may deviate by an angle $\zeta$ from the normal vector \textbf{n}, which is dependent only on geometry of the space curve. Then Eqs.~\eqref{Iout} -- \eqref{Ipi} are modified, after replacing in the integrand $\cos\psi$ by $\cos(\psi-\zeta)$, to
\begin{align}
I_p &=\frac{1}{\pi r^2}\int_0^r \frac {\rho^3}{R^2} \,d \rho  
\int_{\psi}^{2\pi-\psi}\cos^2 (\phi-\zeta) \,d \phi  \notag \\
&=  \frac{(\kappa r)^2}{4\pi}\left(\pi-\psi- \frac 12 \sin 2\psi \cos 2\zeta\right), \label{Ioutz} \\
I_a &=\frac{1}{\pi r^2}\int_0^r \rho \,d \rho 
\int_{-\psi}^{\psi} \left(\kappa \rho \cos(\phi-\zeta) -\epsilon\right)^2 d \phi  \notag \\
 &= \frac{(\kappa r)^2}{4\pi} \left(\psi +\frac 12 \sin 2\psi \cos 2\zeta\right)\notag \\
 &- \frac{4\kappa r\epsilon}{3\pi}\sin \psi \cos \zeta+  \frac{\epsilon^2\psi}{\pi} , \label{Iinz} \\
I (\kappa,\psi) &= I_a+I_p = \frac 14  \left(\kappa r\right)^2 
 -\frac {4\,\epsilon}{3 \pi} \kappa  r \sin\psi \cos \zeta+\frac {\epsilon^2\psi}{\pi} .
 \label{Ipiz}
\end{align}
%

\subsection{Geometry and twist} \label{S22}
 
The twist $\theta$ measures how consecutive cross-sections rotate along the filament. A Janus filament may possess intrinsic torsion $\tau_0$, which is embedded during manufacturing. Then the twist momentum is \cite{Tobias} 
\begin{equation}
J=  \frac{r^2}{6}\theta^2 =\frac{r^2}{6}(\zeta_s + \tau - \tau_0)^2,
\label{Twist}
\end{equation}
where $\tau$ is the geometric torsion of the curve defined in Frenet frame and the index denotes the derivative with respect to the arc length $s$. The intrinsic torsion does not change upon actuation, while $\tau$ and $\zeta$ are dynamic variables defining, alongside curvature, deformation of the filament. The equilibrium values of $\kappa, \, \tau$, and $\zeta$ should be determined while accounting for both the bending and twist, and the problem becomes nonlocal due to the imbedded dependence on changes along the filament.  

It is advantageous to describe three-dimensional (3D) configurations in a coordinate-free form, with the help of the Frenet--Serret equations relating curvature and torsion to changes of the Euclidean position ${\bf x}(s)$ along the centerline:
\begin{equation}
{\bf x}_s= {\bf l},  \;\;\; {\bf l}_s= \kappa {\bf n},  \;\;\;
{\bf n}_s= - \kappa {\bf l} + \tau {\bf b},  \;\;\; {\bf b}_s= -\tau {\bf n},
\label{LFS}  \end{equation}
where ${\bf l},  \;{\bf n},  \;{\bf b}$ are, respectively, the tangent, normal, and binormal unit vectors. Making use of the orthogonality of the vectors forming the Frenet trihedron, the local curvature and torsion are determined, respectively, by taking the scalar product of the second equation \eqref{LFS} with {\bf n} and of the third equation with {\bf b} as 
\begin{equation}
\kappa = {\bf n}\cdot {\bf l}_s,  \qquad \tau = {\bf b}\cdot {\bf n}_s.
\label{kaptau}  \end{equation}

The change of  $\zeta$ along the filament can be found with the help of the rotation vector $\Omega$ which defines the rate of rotation of a local coordinate frame along the filament \cite{Tobias}. The derivative with respect to the arc length of any unit vector \textbf{v} spanning a particular coordinate frame is expressed through the vector $\Omega$  as
 \begin{equation}
\mathbf{v}_s =\Omega \times \mathbf{v}.
\label{omegav}  \end{equation}
Taking the vector product of this relation with \textbf{v} and using the representation of the triple vector product through scalar products yields    
 \begin{equation}
\Omega =\mathbf{v}\times \mathbf{v}_s +(\Omega \cdot\mathbf{v})\mathbf{v}.  
\label{omega}  \end{equation}
In particular, choosing $\mathbf{v}=\mathbf{l}$ and $\mathbf{v}=\mathbf{n}$ and using the Frenet--Serret equations \eqref{LFS} we obtain
 \begin{equation}
 \Omega =(\Omega \cdot\mathbf{l}) \mathbf{l}+\kappa \mathbf{b}=  
(\Omega \cdot\mathbf{n}) \mathbf{n}+\kappa \mathbf{b}+\tau \mathbf{l},
\label{omln} 
\end{equation}
leading to 
 \begin{equation}
 \Omega =\tau \mathbf{l}+\kappa \mathbf{b}, \qquad \Omega \cdot\mathbf{n} =0.
 \label{omln1} 
\end{equation}
Applying Eqs.~\eqref{omegav}, \eqref{omln1} to rotation of $\mathbf{d}={\bf n}\cos\zeta+{\bf b}\sin\zeta$ in the imbedded frame yields
 \begin{align}
 \Omega &= \mathbf{d}\times \mathbf{d}_s +(\Omega \cdot\mathbf{d})\mathbf{d}
 =\kappa \mathbf{d}\sin\zeta+ ({\bf n}\cos\zeta+{\bf b}\sin\zeta) \notag \\
&\times[ \zeta_s({\bf b}\cos\zeta-{\bf n}\sin\zeta) +(\tau {\bf b}-\kappa{\bf l})\cos\zeta- \tau {\bf n}\sin\zeta]  
\notag \\
& \hspace{3cm} = \zeta_s{\bf l} +\kappa{\bf b},
\label{omd} 
\end{align}
 implying $\zeta_s =\tau$.

\section{Open-ended Janus filaments}\label{S3}

\subsection{Design of actuated forms} \label{S31}

A variety of shapes can be obtained already by bending a single Janus filament, provided variable orientations of the driven sector are controlled during manufacturing. The equilibrium shape of a Janus filament with a given position of the driven sector, i.e. direction of the basis vector {\bf d} in each cross-section, can be, in principle, computed by integrating  the Frenet--Serret equations \eqref{LFS}. An open-ended filament can be constrained only by self-intersections. Assuming they are absent, $\bf{n}=\bf{d}$, so that $\zeta=0$. The equilibrium shape should be then untwisted since there are no applied torques, and, according to Eq.~\eqref{Twist}, $\tau=\tau_0$, while the equilibrium curvature is given by Eq.~\eqref{khat}. This uniquely determines the shape, up to spatial translations and rotations or the entire filament.
   
Rather than computing the geometric shape acquired by a specifically constructed Janus filament, we are often interested in \emph{inverse design}, \emph{i.e.} determining the size and orientation of the driven sector in each cross-section along the filament that would generate a desired shape upon actuation. The shape to be attained can be defined either analytically or numerically, in a discretized form, as will be described in detail in Sect.~\ref {S33}. In both cases, the curvature and torsion should be read from the desired shape, and the size and orientation of the driven sector along the filament computed with the help of  Eqs.~\eqref{khat}, {Twist}.

	\begin{figure}[t]
			\begin{tabular}{cc} 
			(a)&(b)\\
			\includegraphics[width=.09\textwidth]{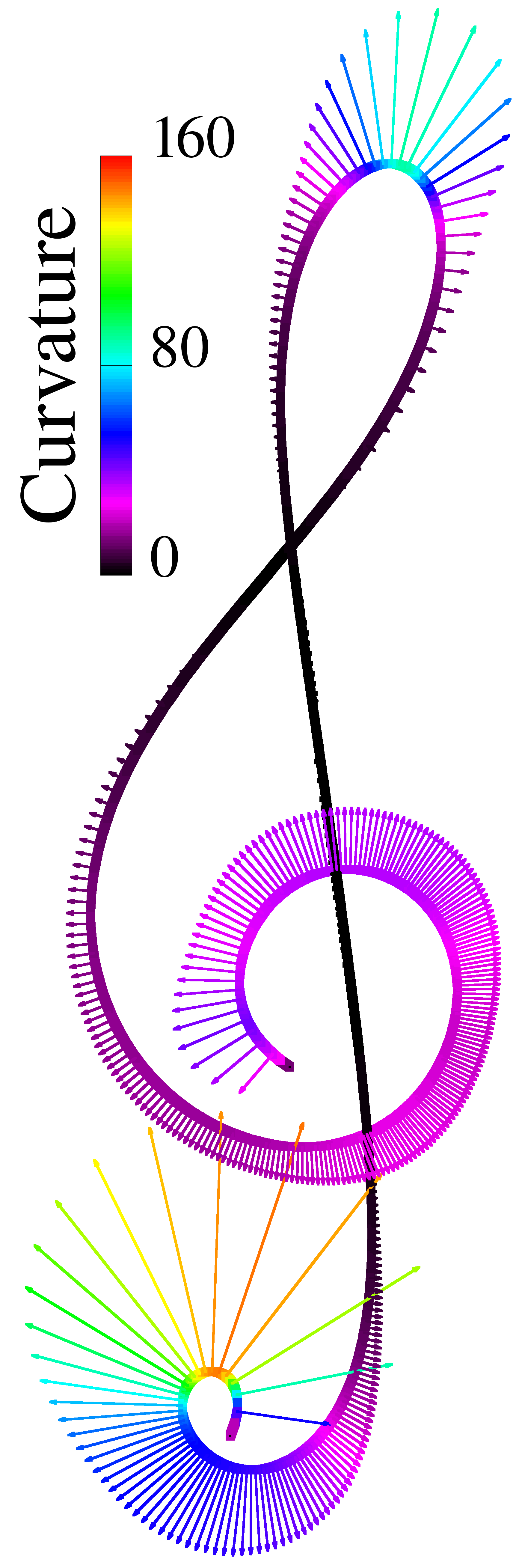}
			&\includegraphics[width=.32\textwidth]{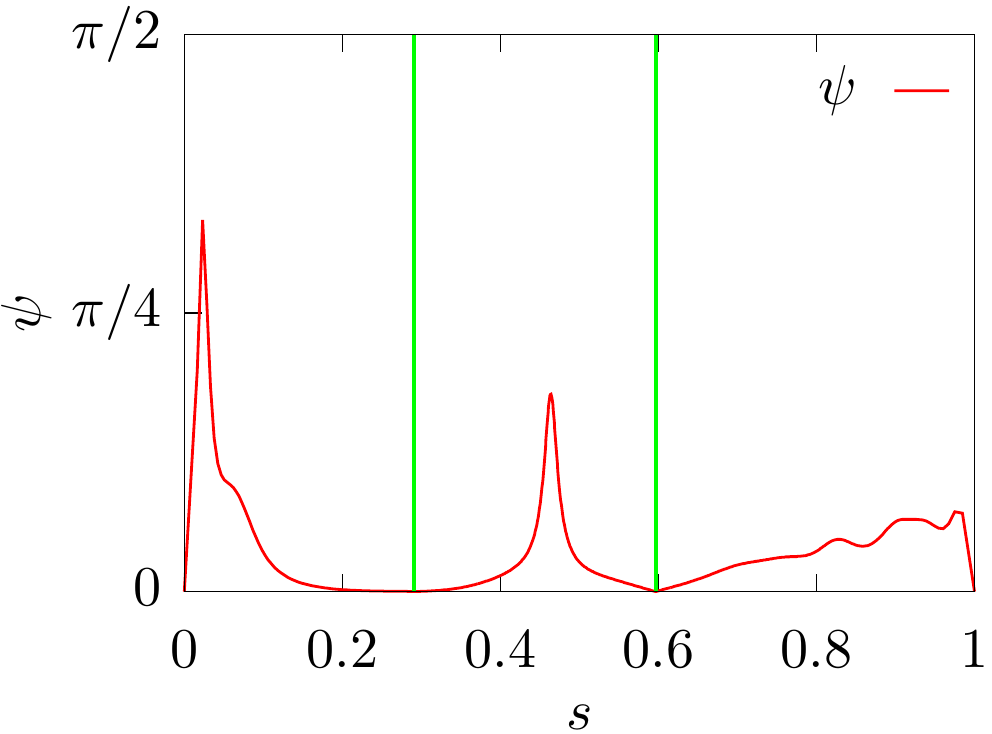}
			\end{tabular}
			\caption{Planar transformation of a filament of unit length. (a) The clef shape with the length of the normals proportional to the local curvature. (b) The angle $\psi$ of the driven sector; the vertical lines correspond to the inflection points. Parameters: $\epsilon=0.2, \, r=10^{-3}$}
			\label{Clef}
\end{figure}  

In the simplest case of two-dimensional (2D) reshaping, the area occupied by the driven component vanishes at inflection points, and the driven sector reappears on the opposite side after this point is passed; these events are marked by vertical lines in Fig.~\ref{Clef} that demonstrates the way of actuating an initially straight filament into a clef-like shape. Out-of-plane bending, which leads to three-dimensional forms, is attained by rotating the driven sector ($\tau_0\neq0$), which can be engineered by rotating coupled extruder streams molding a Janus filament. The variety of forms obtained in this way is limited only by self-intersections of filaments and the magnitude of dilation.

\subsection{Analytical inverse design}\label{S32}
%
\begin{figure}[t]
		\begin{tabular}{ccc} 
		  (a) & (b) & (c)\\
		\includegraphics[width=.13\textwidth]{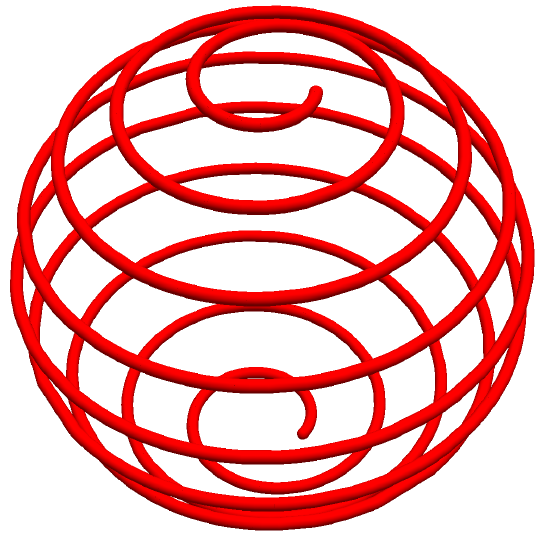}
		&\includegraphics[width=.15\textwidth]{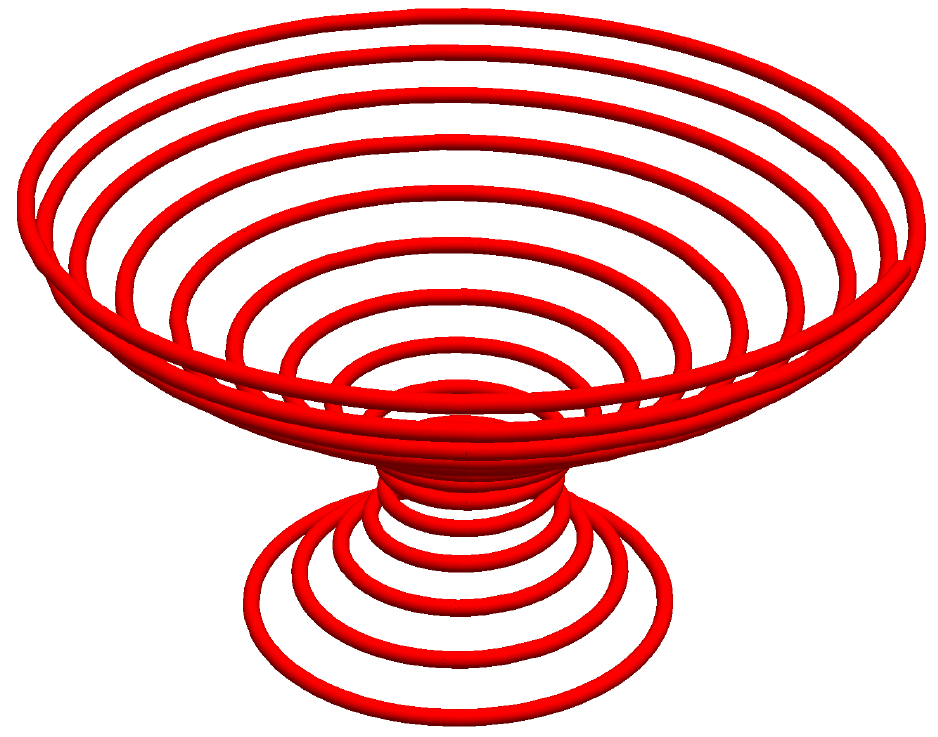}
		&\includegraphics[width=.16\textwidth]{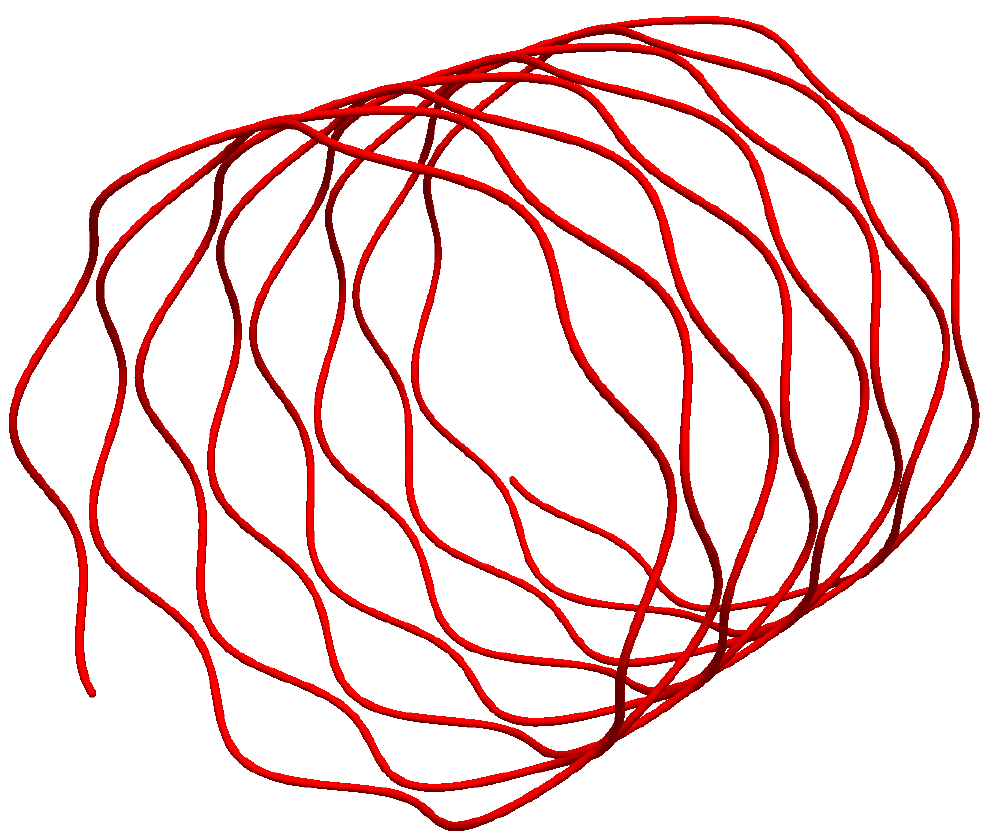}\\
		  (d) & (e) & (f)\\
	\includegraphics[width=.15\textwidth]{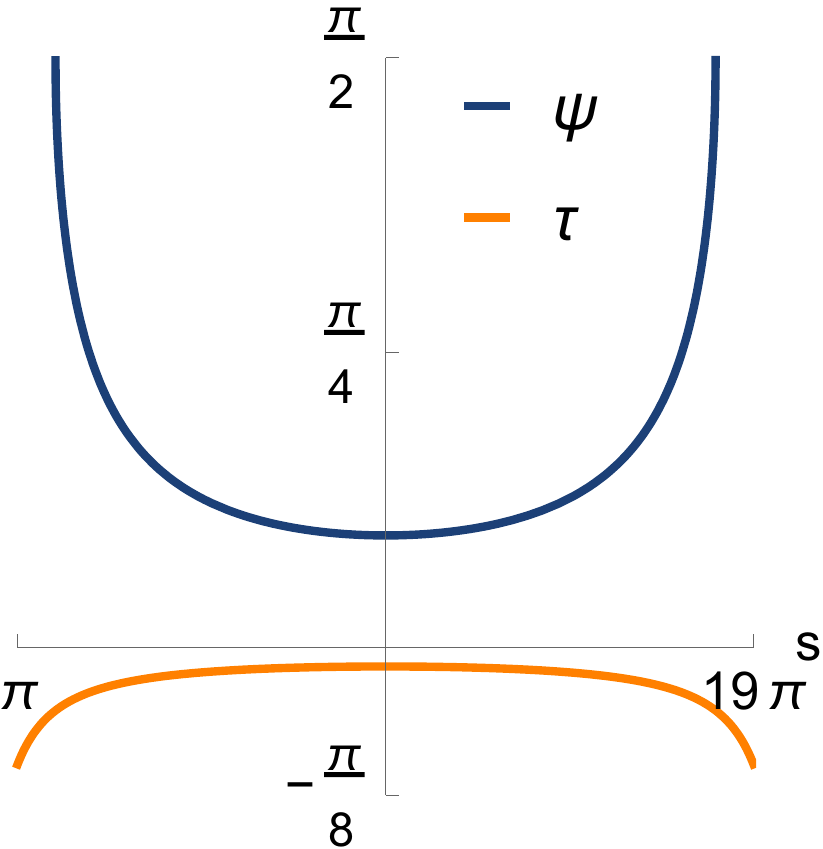}
		&\includegraphics[width=.15\textwidth]{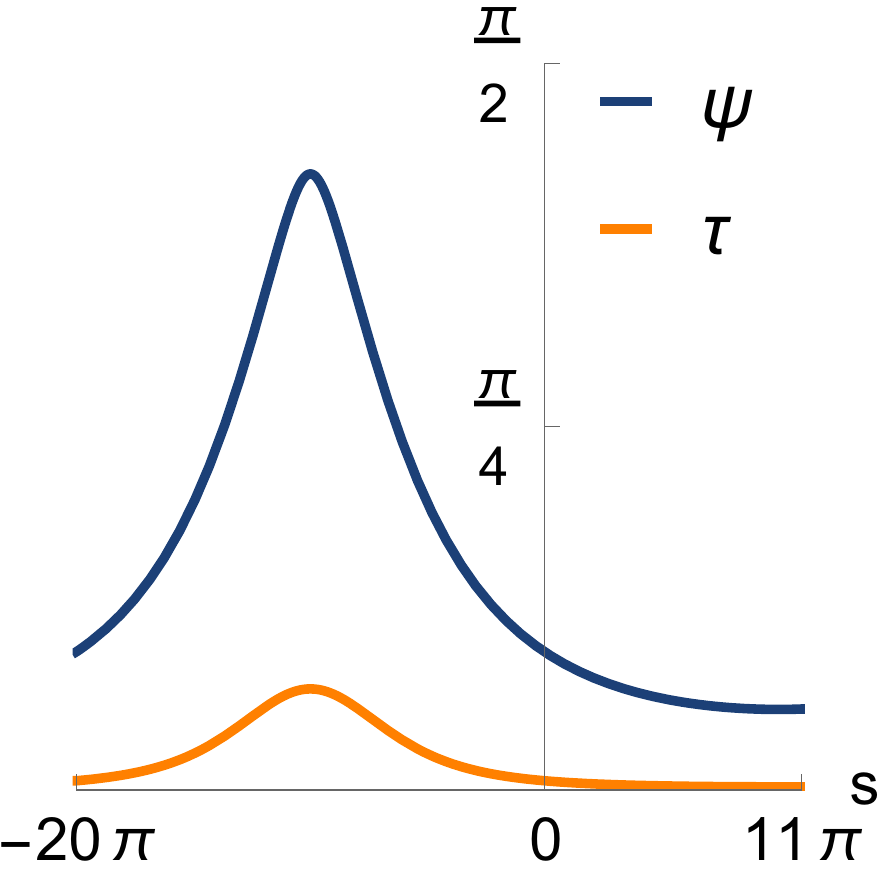}
		&\includegraphics[width=.16\textwidth]{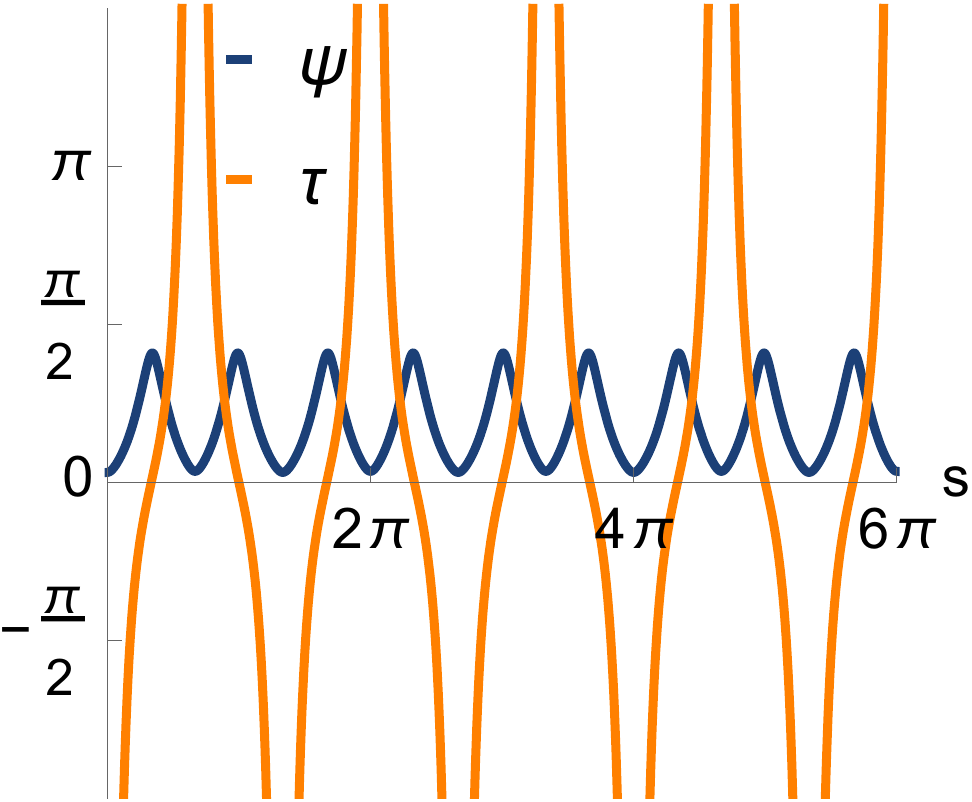}
		\end{tabular}
		\caption{(a--c): The various 3D shapes made of a single preprogrammed filament. (d--f): The size of the driven sector and torsion as functions of the arc length $s$}
		\label{Examples}
\end{figure}  

While inverse design of 2D shapes is straightforward, there are multiple ways to obtain even a simple 3D shape delineated by filaments. Since we are interested in neither an optimal layout nor a uniform surface covering, we attempt here to reproduce a desired 3D shape by covering the target surface by a single filament, and only require the filament to remain smooth with no sharp kinks and self-intersections. It is natural therefore to describe a slender deformable filament as a space curve with the coordinates ${\bf x}(s)$ parametrized by the arc length $s$, and determine its local properties with the help of the Frenet frame spanned by the tangent, normal, and binormal unit vectors, respectively, ${\bf l},\,{\bf n},\,{\bf b}$. 

If the desired shape can be expressed analytically as a continuous curve with the Euclidean position defined as a vector-function ${\bf x}(s)$ of the arc length $s$, the curvature and torsion are computed straightforwardly with the help of Eqs.~\eqref{kaptau}. The simplest analytically treatable 3D example is a helix $x(t)=\{\cos t, \, \sin t, \,at\}$ parametrized by a variable $t$ related to the arc length as $s=t \sqrt{1+a^2}$. Then Eq.~\eqref{kaptau} yields constant curvature and torsion
\begin{equation}
\kappa =1/\sqrt{ 1+a^2},  \qquad \tau = a/\sqrt{ 1+a^2}.
\label{kaptau1}  \end{equation}
The size of the driven sector is characterized by the angle $\psi$, which is required to deform an originally straight-line filament into the helix with the unit radius and pitch $a$, is defined then by Eq.~\eqref{khat} with $\widehat{\kappa}$ given by the first Eq.~\eqref{kaptau1}, while the second Eq.~\eqref{kaptau1} defines rotation of the driven sector alongside the filament.  

Less trivial examples are provided by the three shapes depicted in Fig.~\ref{Examples}(a--c) obtained by parametrizing appropriate spatial curves. For the spherical shape in the left-hand panel, the Euclidean positions are defined as 
\begin{equation}
{\bf x}(t)=\{\sin a t \cos t ,\;\sin a t \sin t ,\;\cos a t \}, 
\end{equation}
where $a=0.05$, $t\in[\pi;19\pi]$. For the cup-like shape in the left panel, we have 
\begin{equation}
{\bf x}(t)=\{(b\sin a t +1)\cos t ,\;(b\sin a t +1)\sin t ,\;c t\},
\end{equation}
where $a=0.05$, $b=0.7$, $c=0.02$, $t\in[-20\pi;11\pi]$, and for the wavy cylinder in the right-hand panel, 
\begin{equation}
{\bf x}(t)=\{\cos a t ,\;\sin a t ,\;b t+c\sin d t \}
\end{equation}
 with $a=0.2$, $b=0.01$, $c=0.1$, $d=1.5$, $t\in[0;80\pi]$. In these examples, both the curvature and torsion vary along the length of the filament; their dependence on the variable $t$ are defined (after normalizing $t$ to the arc length) by the same Eq.~\eqref{kaptau}, though the respective analytical expressions are far more involved than Eq.~\eqref{kaptau1}. The resulting change of $\psi$ and rotation of the orientation angle of the basis over the length of the filament are shown in Fig.~\ref{Examples}(d--f). Take note that in Fig.~\ref{Examples}f the torsion diverges at inflection points where the curvature vanishes. As in the 2D example in Fig.~\ref{Clef}, the vanishing driven sector reappears then at a different location.

\begin{figure}[t]
			\begin{tabular}{cc} 
			(a)&(b)\\ 
			\includegraphics[width=.22\textwidth]{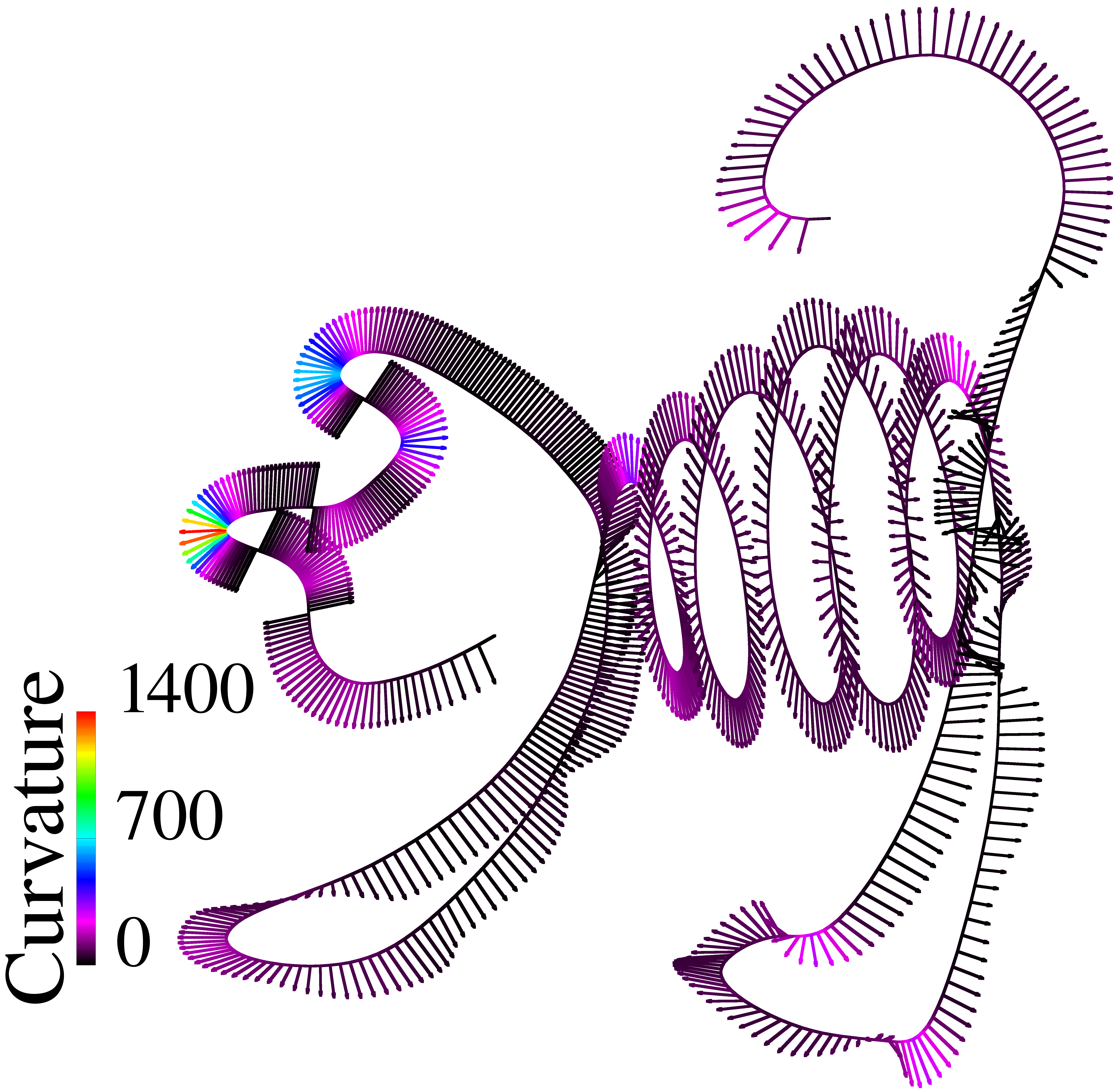}
			&\includegraphics[width=.22\textwidth]{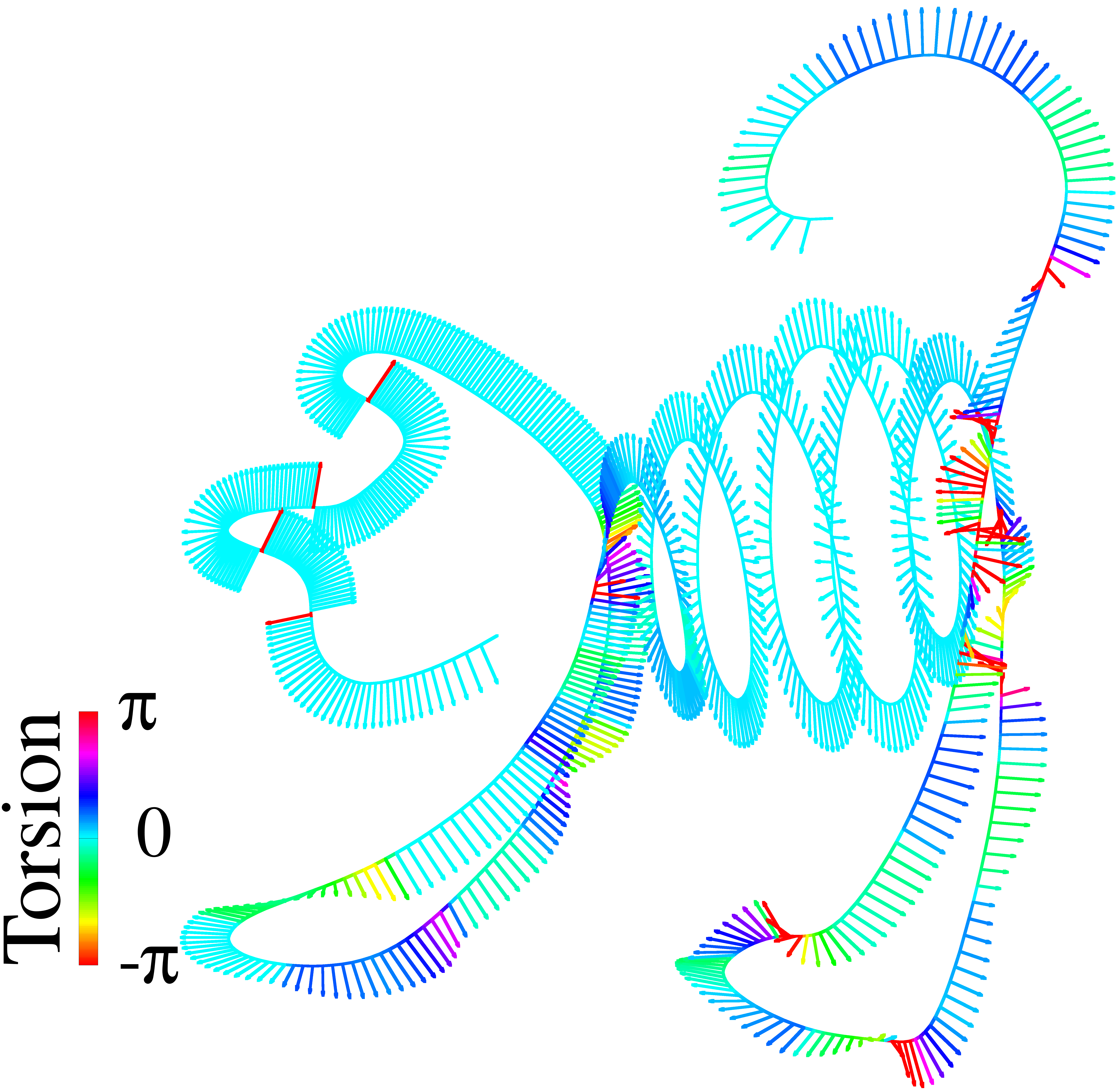}
						\end{tabular}
			\centerline{(c)}\\
			\includegraphics[width=.43\textwidth]{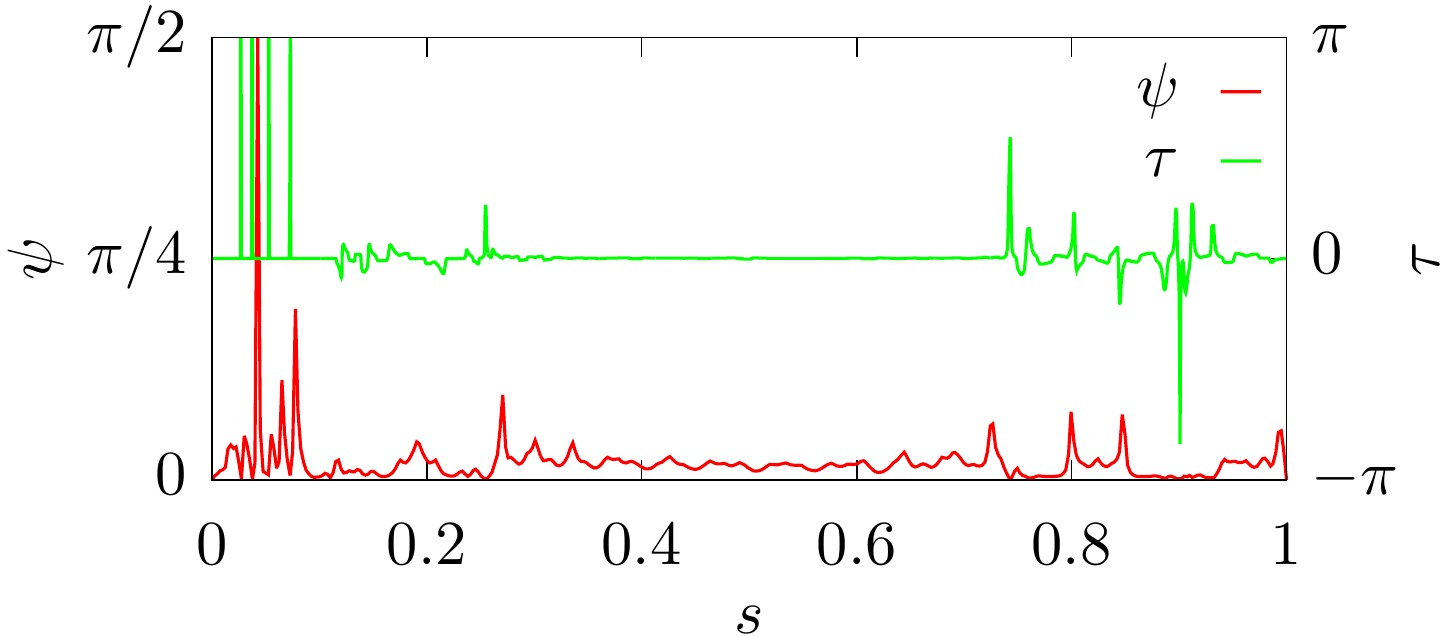}	
			\caption{(a,b): A filament of unit length reshaping to a cat sculpture. The normals are colored according to the local curvature (a) and twist (b). (c): The size of the driven sector and torsion as functions of the arc length $s$Parameters: $\epsilon=0.2, \, r=10^{-4}$}
			\label{Cat3D}\end{figure}  
%
		                
\subsection{Numerical inverse design}\label{S33}

If the desired shape is given numerically, the local curvature and torsion are obtained by discretizing the centerline of a filament as a chain of nodes. Then the local curvature can be conveniently calculated at each $i$th node as an approximant $\kappa_i = 2 \alpha_i/(l_{i-1}+l_i)$, where $\alpha_i \ll 1$ is the angle between two links adjacent at the $i$th node and $l_{i-1}$, $l_i$ are the distances between the points $i-1$, $i$ and $i$, $i+1$, respectively. This approximation is valid as long as the local curvature radius is large compared to the distance between discretization points. The approximated torsion $\tau$ is computed on an $i$th link  as $\tau_i = -\beta_i/l_i$, where $\beta_i$ is the angle between projections of normals at the points $i-1$ and $i+1$  onto the normal plane of the $i$th link, with the sign defined by the counter-clockwise rotation convention. 
Once the geometric data are known, the size and orientation of the driven sector are computed as outlined in Sect.~\ref{S31}. 

An example of inverse design applied to a rather complex shape imitating a wire sculpture \cite{wiki} is shown in Fig.~\ref{Cat3D}. The coloring or shading of the normals in these pictures is proportional to the local curvature (a) and torsion (b). The distribution of $\psi$ and $\tau$ along the filament is  shown in Fig.~\ref{Cat3D}c.

\section{Deformation of a closed Janus filament}\label{S4}
\subsection{Virtual displacements}\label{S41}

In a closed Janus filament with a constant area and orientation of the driven sector, instabilities are apt to arise due to an interplay between bending and twist.  In an untwisted circular ring with a radius $1/\kappa_0$, originally flat and torsionless, the basis $\mathbf{d}$ is parallel to the Frenet normal; this, however is apt to change when the ring relaxes to an equilibrium curvature $\widehat{\kappa}<\kappa_0$, which, due to a constrained geometry of a close ring, necessitates misalignment of the two vectors and the emergence of non-zero torsion. 

The change of energy caused by small deformations can be computed by testing effects of displacing of the centerline defined in a parametric form ${\bf x}(s)$ by increments $u,v,w$ along, respectively, $\bf{l}, \bf{n}, \bf{b}$, so that  
\begin{equation}
\delta \mathbf{x} = u{ \bf l}+v { \bf n}+w{ \bf b}.
\label{Lvel}  \end{equation}
The tangential displacement $u$ is a gauge variable that expresses a  reparametrization of a displaced curve. For a nonextensible filament, perturbing the first Frenet--Serret equation \eqref{LFS} and replacing the derivatives with respect to $s$ with the help of other equations, yields the increment of {\bf l}: 
\begin{equation}
\delta{\bf l} =  \delta{\bf x}_s = (u_s  - \kappa v) {\bf l}  +   (v_s + \kappa u - \tau w) {\bf n} + (w_s + \tau v) {\bf b}.  
\label{Llt}  \end{equation}
Since {\bf l} is a unit vector, the projection $\delta\mathbf{l \cdot l}$ should vanish. This yields the relation between $u$ and $v$ in the isometric gauge
\begin{equation}
u_s = \kappa v. 
\label{Lgt}  \end{equation}
Equation (\ref{Llt}) thus reduces to
\begin{equation}
\delta{\bf l} = V {\bf n} + W{\bf b}; \;\;\; 
V =v_s + \kappa u - \tau w, \;\;\; W = w_s + \tau v.  
\label{Llt1}  \end{equation}
In the same manner, we obtain the increment of {\bf n} from the second Frenet--Serret equation with the help of Eqs.~(\ref{Lgt}) and (\ref{Llt1}) : 
{\samepage \begin{eqnarray}
\kappa\, \delta{\bf n} &= & - \delta \kappa\,{\bf n}
+ \partial_s (V {\bf n} + W{\bf b})  = - V {\bf l}+  (W_s +  \tau V) {\bf b} \nonumber \\
&+&  (-\delta \kappa  -\kappa u_s + \kappa^2 v +V_s - \tau W) {\bf n}.  
\label{Lnt}  \end{eqnarray}}
Again, requiring the projection $\delta\mathbf{n \cdot n}$ to vanish and using Eq.~(\ref{Lgt}) gives the increment of the curvature:
\begin{equation}
\delta\kappa  = V_s - \tau W. 
\label{Lkt}  \end{equation}
The perturbation of {\bf n} reduces therefore to
\begin{equation}
\delta{\bf n} = - V {\bf l} + U{\bf b}; \qquad
U = \kappa^{-1}( W_s +  \tau V).  
\label{Lnt1}  \end{equation}
The increment of the torsion $\tau$ is obtained from the third Frenet--Serret equation after using Eq.~(\ref{Lgt}) and requiring the projection $\delta{\bf b \cdot b}$ to vanish:
\begin{equation}
\delta\tau  =  \kappa W + U_s. 
\label{Ltt}  \end{equation}
%
 
\subsection{Stability of an untwisted ring}\label{S42}

The linear stability of a ring can be studied by testing the effect of infinitesimal harmonic perturbations
 \begin{align}
& v=\widetilde{v}\E^{\I k s}, \quad w=\widetilde{w}\E^{\I k s},\quad \quad \tau=\widetilde{\tau}\E^{\I k s}, \notag \\
& u =\widetilde{u}\E^{\I k s}= -\frac{\I \kappa_0}{k}\widetilde{v}, \quad
  \zeta =\widetilde{\zeta}\E^{\I k s}= -\frac{\I}{k}\widetilde{\tau}\E^{\I k s},
\label{linu} 
\end{align}
where the last two relations follow from Eqs.~(\ref{Lgt}), (\ref{omd}) and the wavenumber $k=n \kappa_0$ should be an integer multiple of $\kappa_0$. 

The change of energy $\delta\mathcal{F}$ is quadratic in the perturbation amplitudes, and the basic configuration is unstable when $\delta\mathcal{F}<0$. For the purpose of linear stability analysis, we need to compute the lowest order (quadratic) resonant terms in Eq.~\eqref {Ft} to determine whether the energy decreases upon virtual infinitesimal displacements of the filament and rotations of the basis defined by Eqs.~\eqref{linu} and their complex conjugates $\widetilde{v}^*\E^{-\I k s}$ \emph{etc}. The curvature is expressed as  
 \begin{equation}
 \kappa=\kappa_0+\widetilde{\kappa}\E^{\I k s}+\widetilde{\kappa}^*\E^{-\I k s} + \overline{\kappa},
 \label{lint} \end{equation}
where $ \overline{\kappa}$ denotes constant terms quadratic in perturbation amplitudes. The relations between the curvature and torsion and virtual displacements follow from Eqs.~(\ref{Lkt}), (\ref{Ltt}). Collecting the first-order terms only yields
\begin{equation}
\widetilde{\kappa}=- (k^2-\kappa_0^2) \widetilde{v}, \qquad
\widetilde{\tau}=- \frac{\I k} {\kappa_0} (k^2-\kappa_0^2) \widetilde{w}.
\label{kapv}  
\end{equation}
Collecting quadratic terms in Eq.~\eqref {Lkt} yields the second-order resonant term 
 \begin{equation}
\overline{\kappa}  =  2 (k^2/\kappa_0) (k^2 - \kappa_0^2 )|\widetilde{w}|^2. 
\label{evolk2}
\end{equation}

Eliminating the extension coefficient $\epsilon$ with the help of Eq.~\eqref{khat}, the variable part of the bending moment can be expressed through the equilibrium curvature $\widehat{\kappa}$: 
\begin{equation}
I  =  \frac 14 r^2 \kappa \left( \kappa  -  2 \widehat{\kappa}\cos\zeta \right).
 \label{Fdel}
\end{equation}
The twist moment is given by Eq.~\eqref{Twist} with $\tau_0=0$ is, in view of Eq.~\eqref {omd}, $J=\frac 23 \tau^2$. The relevant non-oscillating second order terms in the integrand of Eq.~\eqref{Fe1} are computed using Eqs.~\eqref {kapv}, \eqref {evolk2} and replacing $k=n \kappa_0, \, \overline{\kappa}=q\kappa_0$ as 
 \begin{align}
{\mathcal{I}} &= \frac 12 r^2 \left[ |\widetilde{\kappa}|^2+ \overline{\kappa}( \kappa_0-\widehat{\kappa})+\kappa_0 \widehat{\kappa}|\widetilde{\zeta}|^2\right] + \frac 23 |\widetilde{\tau}|^2 \notag \\
&=   r^2 \kappa_0^4 \left[ \frac 12 (n^2-1)^2|\widetilde{v}|^2 
  + f(n^2) |\widetilde{w}|^2\right],  \label{Ft} \\
f(n) &=  (n^2-1)^2\left( n^2 \frac{1-q}{n^2-1}+\frac q2 +\frac 23n^2 \right)
 \label{fk}
\end{align}
Notably, this expression does not involve the size of the driven sector, which only affects the equilibrium curvature $\widehat{\kappa}$.

Normal displacements affect only the perturbation of $\kappa$, and the filament is always stable to normal displacements. Instability to binormal displacements is determined by the function $f(k)$ in Eq.~\eqref{fk}. The instability sets on at $q>q_c$, where $q_c$ is obtained by solving $f(n)=0$: 
\begin{equation}
 q_c = \frac 23 \frac{ n^2 (2+ n^2)}{2+ n^2}. 
  \label{Ftq}
\end{equation}
The numerical values are $q_c  = 3.2$ at $n=2$, $q_c  = 6.6$ at $n=3$, $q_c \approx 11.3$ at $n=4$. These values differ from those reported in Ref.~\cite{pre18} but not in a qualitative way. A twisted ring can be proven to be absolutely unstable to infinitesimal perturbations, in accordance with  Ref.~\cite{pre18}.

\begin{figure}[t]
\centering
  \includegraphics[width=.40\textwidth]{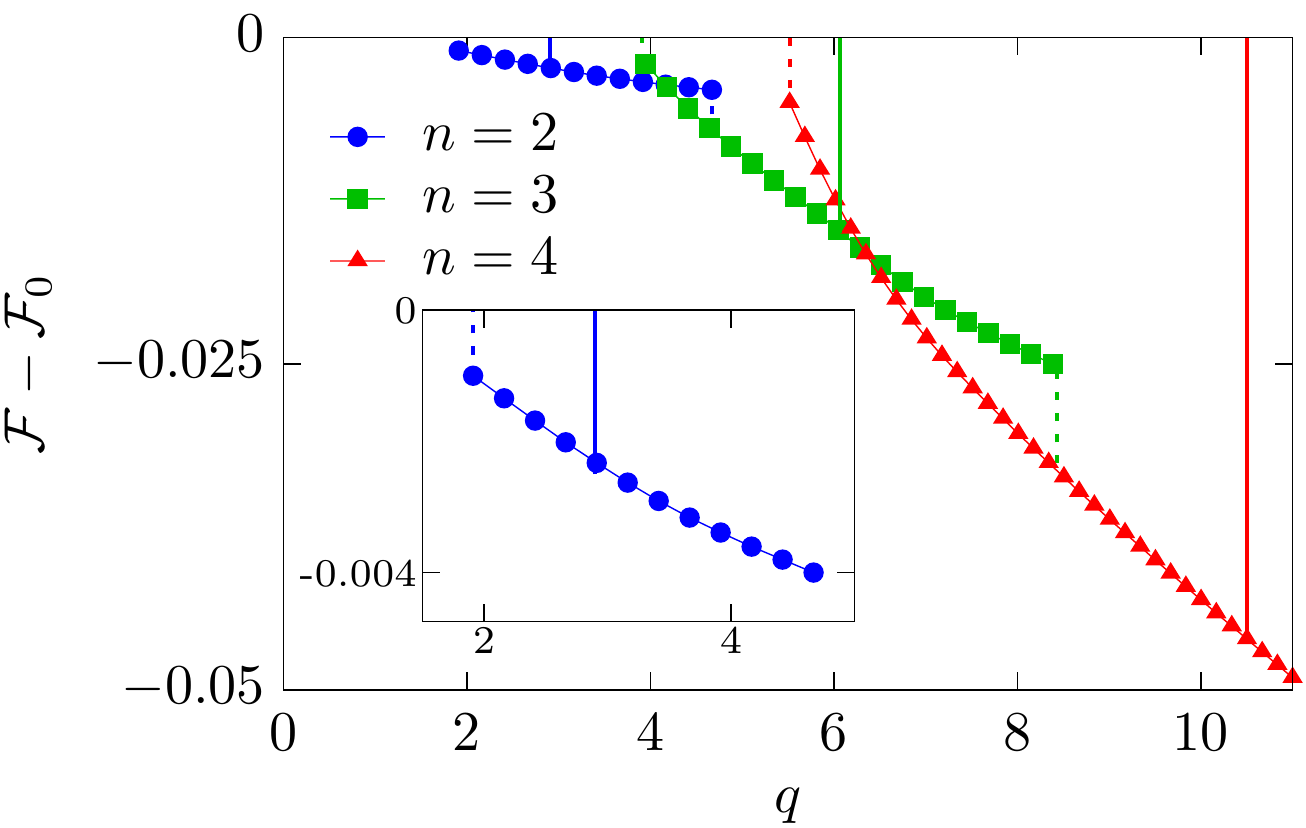}\\
\caption{ Elastic instability transitions in a closed Janus filament under actuation. The overall energy gain by reshaping of filament to stable and metastable configurations at different dimensionless parameter $q$ and initial perturbation wavelengths. Instability thresholds found in stability analysis are plotted by solid lines. Dashed lines show transitions to a higher looping number.}  
\label{Jring} \end{figure}
%

\subsection{Large-amplitude deformations}\label{S43}

Large-amplitude deformations are computed by numerical simulations based on approximating the filament centerline as a discrete curve with material frames defined at nodes. Equilibrium configurations can be attained by solving the energy minimization problem. The local curvature and torsion are computed as described in Sect.~\ref{S33}, while angle $\zeta$ in Eq.~\eqref{Twist} depends on mutual orientation of the Frenet frame and the material basis defined in the reference state. 
The energy minima are located using the gradient descent method by computing energy gradients due to imposed small virtual local perturbations of each degree of freedom, i.e. by three spatial displacements and rotation of the material frame. After each simultaneous update of node positions and orientations, calculation of gradients and overall energy is repeated till the energy gain vanishes.

We found that non-planar configurations for each looping number always exist below the instability thresholds $q_c$, so that the instabilities analyzed above are always \emph{subcritical}. The overall energy of the configurations is plotted in Fig.~\ref{Jring} as a function of the dimensionless parameter $q$.
The results indicate multistability with coexisting metastable and absolutely stable configurations with different looping numbers. The general tendency seen in Fig.~\ref{Jring} indicated transition to larger looping numbers with increasing $q$.  Relaxation to a lowest energy configuration is commonly restricted by self-intersections of a filament, so that large-amplitude perturbations are necessary to further decrease the energy while avoiding self-contact.    

\section{Reshaping of an elementary frame}\label{S5}
\subsection{Geometry and energy of an elementary frame}\label{S51}

More possibilities arise when Janus filaments are incorporated in two-dimensional woven structures. Textiles assembled from separate filaments allow for more freedom compared to continuous sheets, as they are not constrained by the condition of incompressibility. The filaments in textiles are relatively disperse and may be able to slide, bend and change distance between them. Stitching at edges ensures the integrity of the fabric and, alongside interlacing, prevents self-intersections.  

\begin{figure}[b]
 \begin{tabular}{ccc}
 	(a)&(b)&(c)\\
 \includegraphics[width=.16\textwidth]{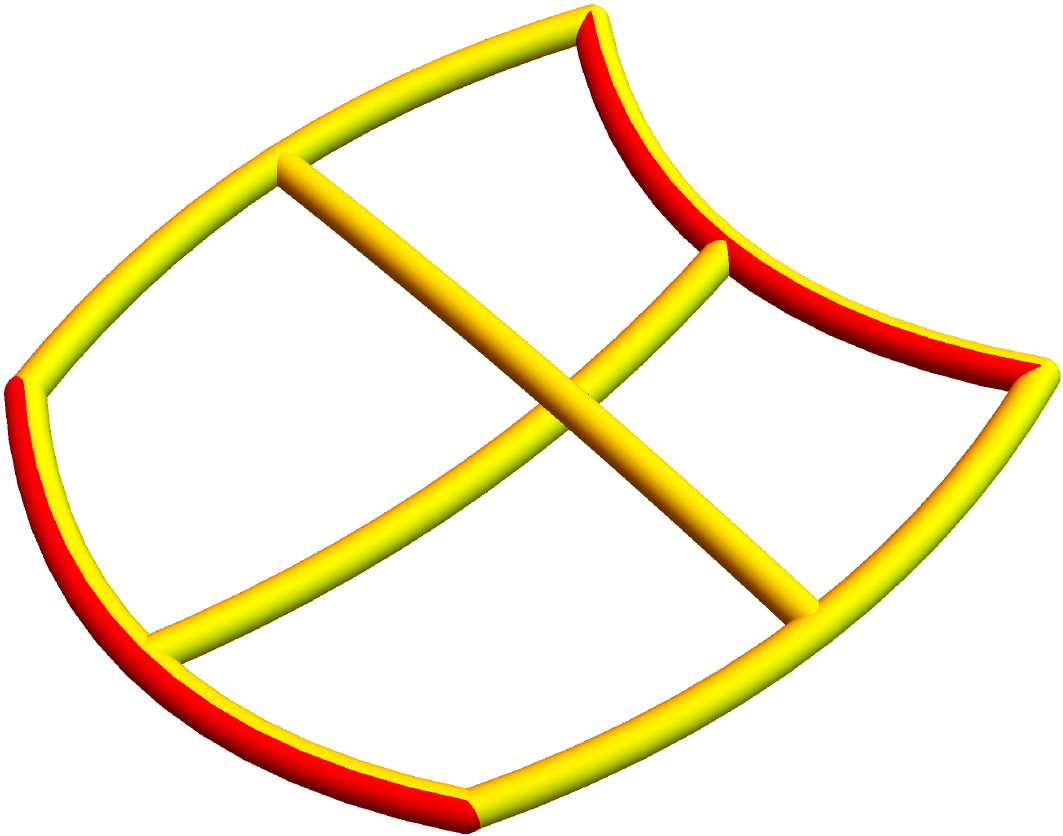}
&	\includegraphics[width=.155\textwidth]{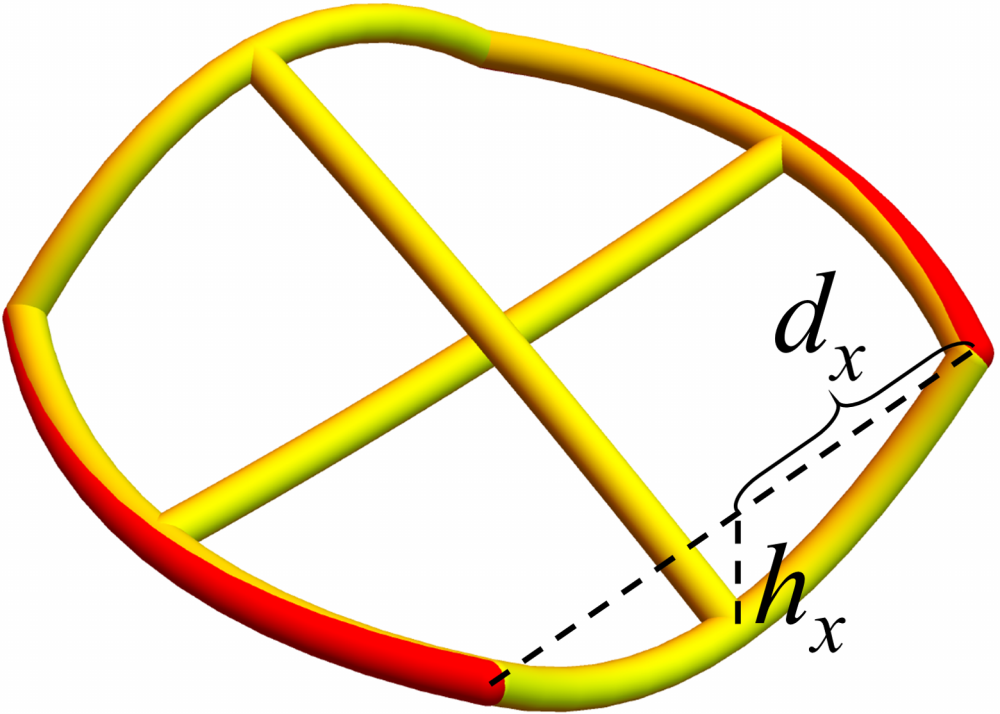}
  &\includegraphics[width=.145\textwidth]{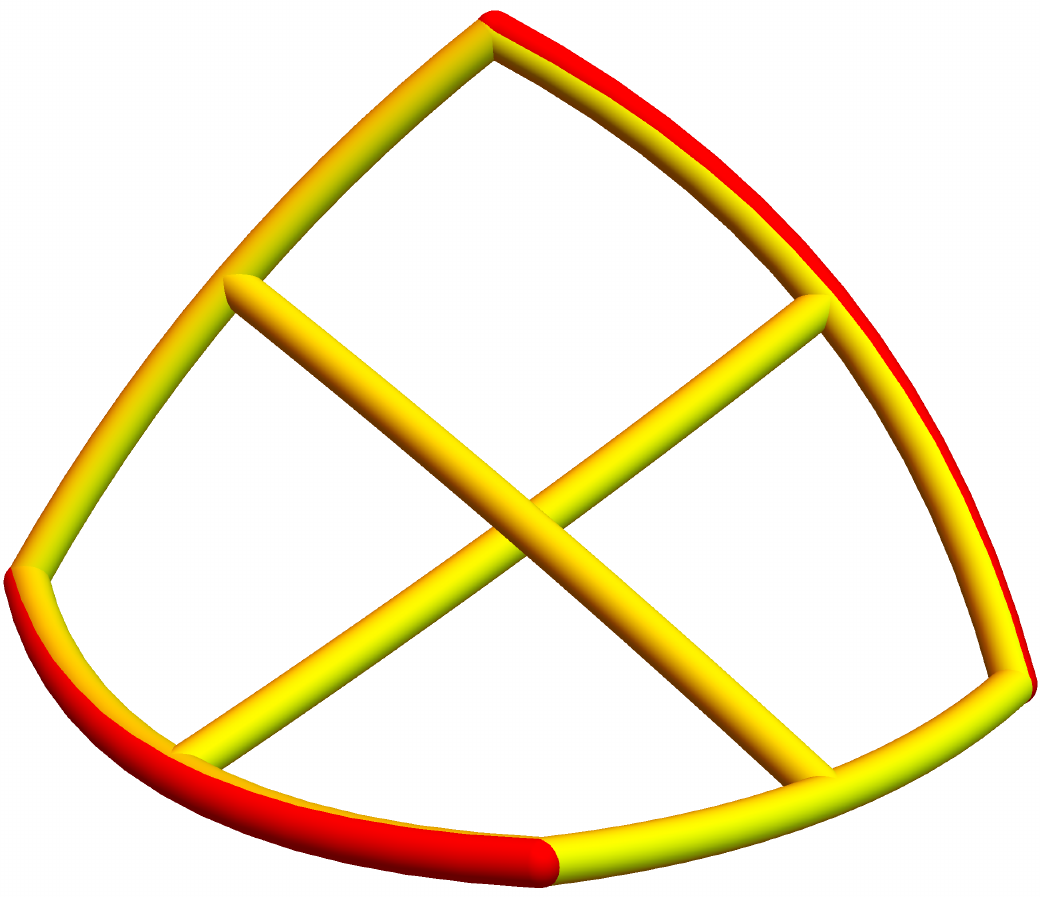}
 \end{tabular}
\caption{Representative configurations of elementary frames with antisymmetric actuation (a) and symmetric (b,c) actuation}  
\label{frame} 
\end{figure}

As a model example allowing for analytical treatment, we consider a minimal piece of textile containing two Janus filaments with half a cross-section occupied by an actuated material. The elementary structures include two Janus filaments and one passive filament placed in between, which are oriented before actuation along the $y$ axis, and three passive filaments positioned in the orthogonal $x$ direction and connecting the ends and centers of the Janus filaments. All filaments are originally straight and have the same length $L$ in the reference state. The filaments are allowed in this model example to freely change connection angles at their ends.
At the same time, relative rotation of the framing filaments cause twist of the two central filaments that has not been accounted for in Ref.~\cite{textile}. Although the integrity of this minimal structure is not supported by weaving, we assume that the two passive filaments intersecting at the central node remain in contact. Some typical shapes obtained upon actuation are shown in Fig.~\ref{frame}.  

The energy of an elementary frame is dependent on the curvatures $\kappa_i$ of the driven, parallel passive, and perpendicular framing and central filaments (marked by the indices $i \to (a,y,x,c)$, respectively) and twists $\theta_c, \theta_y$. Since curvatures of each filament are constant along their length, it is convenient to work with the respective curvature radii $R_i=1/\kappa_i$. The general expression for energy per unit volume, scaled by $E r^4/2$, where the radii $r$ and Young moduli $E$ of driven and passive filaments are assumed to be equal, can be written as 
\begin{align}
\mathcal{F}=& \frac {1}{4} \left[ R_y^{-2} + R_c^{-2} + 2R_x^{-2} + 2I(1/R_a)  \right]
+ \frac {1}{6} \left[\theta_y^2+ \theta_c^2 \right].
 \label{engen}
\end{align}
The bending moment $I(1/R_a) =I(\kappa_a)$ of driven filaments is given by Eq.~\eqref{Ipiz} where we set $\psi=\pi/2$ to maximize the curvature attained upon actuation. As usual, we neglect here a small correction to the radius of driven filaments upon actuation, $1/\sqrt{1+\epsilon/2}$  for LCE or $1+\epsilon/2$ for the case of isotropic expansion. 

To detect the dependence of the curvatures of passive filaments on $R_a$ and $\epsilon$, it is convenient to define arc angles $\phi_i=L_i/R_i$ with the lengths of all passive filaments equal to $L$ and $L_a=L(1+\epsilon/2)$. We will also need the chord half-lengths (half-distances between the ends of bent filaments) $d_i=R_i\sin \phi_i/2$ and the heights of the segments (sagitta) $h_i =R_i(1-\cos \phi_i/2)$. Independently of orientation of driven filaments, the character of deformations is defined by the sign of the difference 
\begin{equation}
\delta =d_a - d_y= \frac 34 \frac{\pi r}{\epsilon} \sin \left[\frac 34 \frac{\epsilon L}{\pi r}
 \left(1 + \frac \epsilon 2\right)\right]-1. 
 \label{chord}
\end{equation}
This difference grows at very small extensions and then decreases and becomes negative above the critical elongation $\epsilon_c=2 R_a \arcsin (L/R_a) - 2$.

\subsection{Antisymmetric actuation}\label{S52}

Consider first the case when both Janus filaments are oriented in the same direction and bend in the frame plane. At $\epsilon<\epsilon_c, \, \delta >0$, elongation of driven filaments upon actuation overcompensates the decrease of $d_a$ caused by bending. A flat configuration is possible under these conditions when the perpendicular framing filaments bend inward with the curvatures $1/R_x$ satisfying $d_a-d_y = h_x$. This deformation reduces the half-distance between the driven filaments to $d_x <L/2$, and therefore the central perpendicular filament must also bend with the same curvature radius $R_x$ to preserve its length, while the parallel passive filament remains straight. The bending should be in-plane (in either direction) if preserving the contact with the rectilinear parallel passive filament at the central node is required, and lifting the central node off-plane would require extending the framing filaments.  
\begin{figure}[b]
\begin{tabular}{cc}
		(a)&(b)\\
 \includegraphics[width=.235\textwidth]{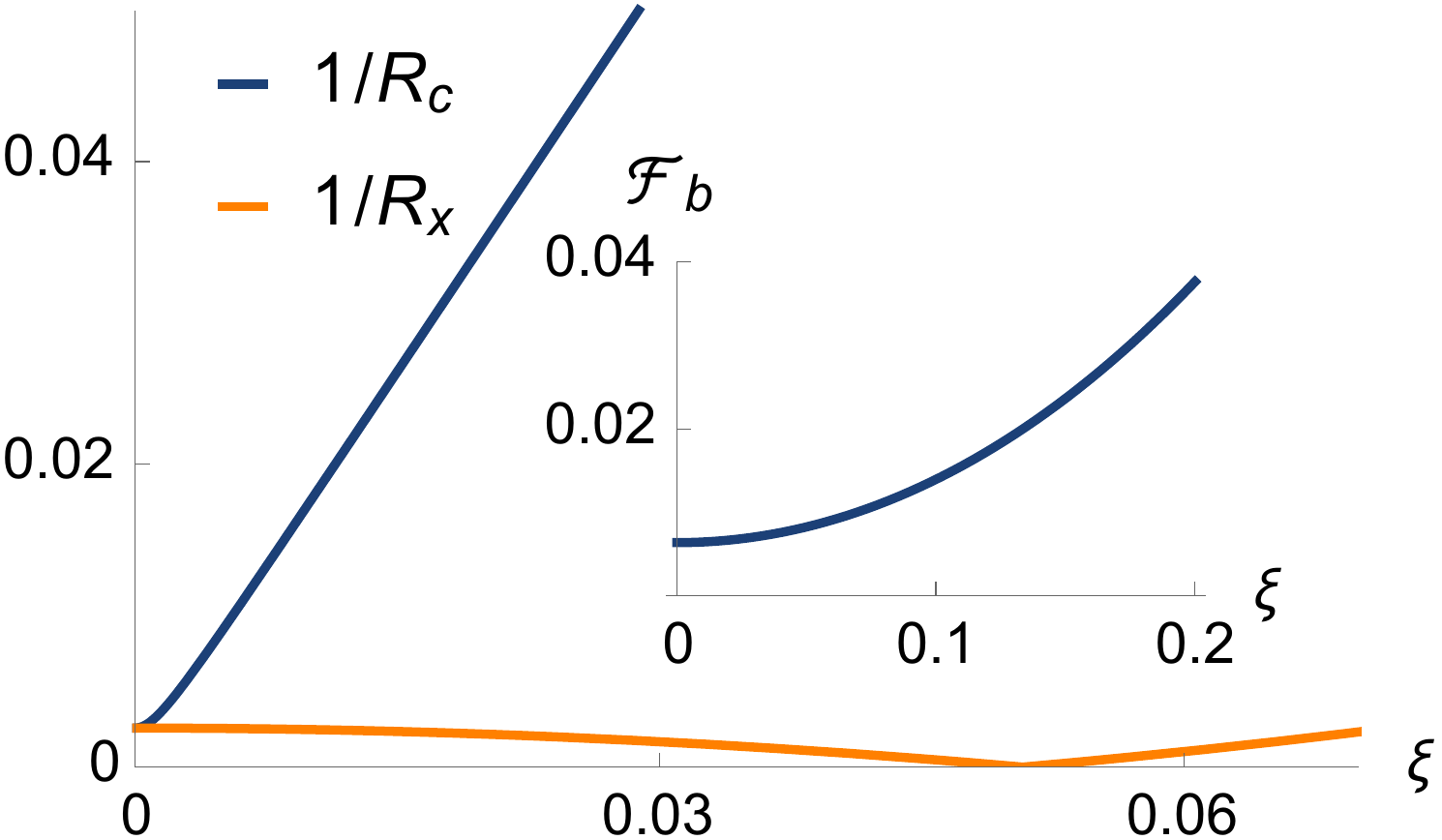}
 &\includegraphics[width=.235\textwidth]{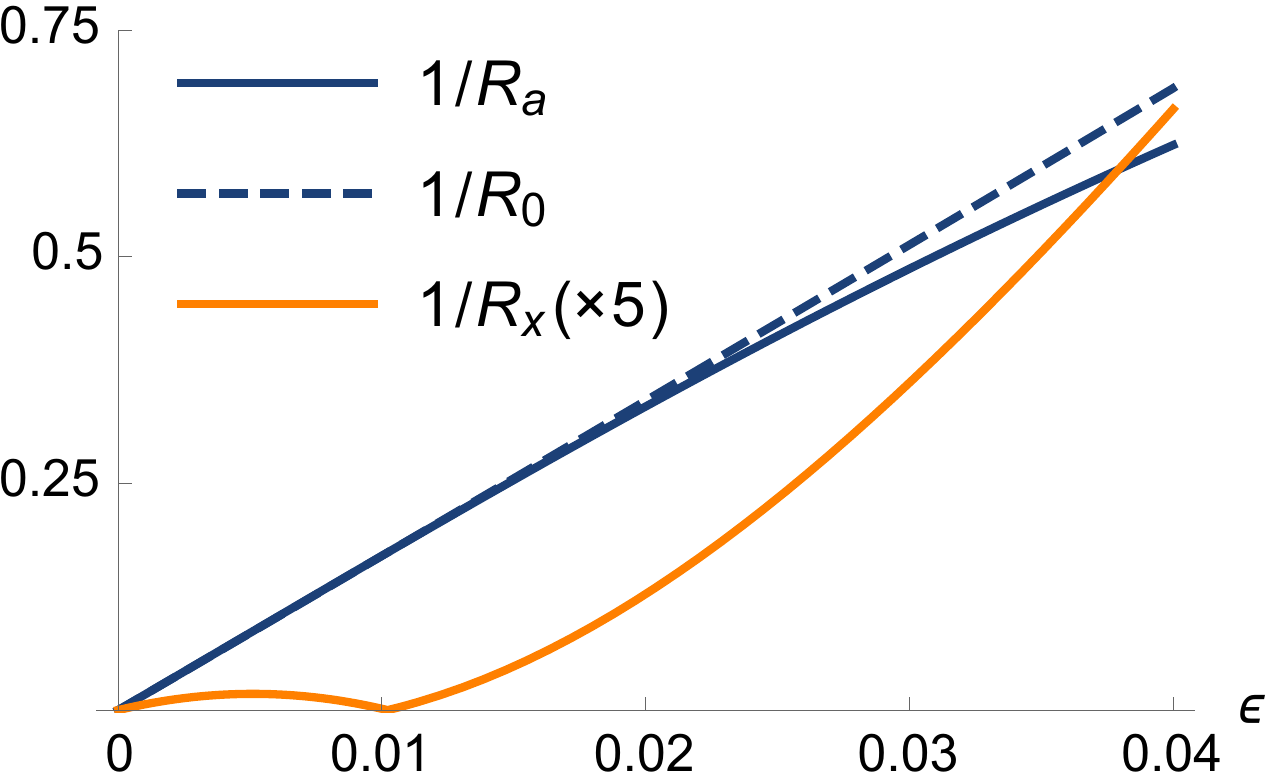}\\
 \end{tabular}
\caption{Antisymmetric bending of the frame made of filaments with $r=0.05$ and  $L=2$.  (a) Dependence of the curvature of perpendicular filaments and the bending energy on rotation of driven filaments about midpoint at $\epsilon=0.005$. (b) The dependence of the curvatures of driven and passive filaments in the equilibrium configuration on the extension coefficient $\epsilon$. }  
\label{AntiSymRaRxEps} 
\end{figure}

The energy of this configuration is
\begin{equation}
\mathcal{F}= \frac 34 R_x^{-2} + \chi I(1/R_a) 
 \label{enflat}
\end{equation}
The energy minimum is determined by trade-off between the energies of perpendicular passive and the driven filaments, leading to a deviation of the latter's curvature from the optimal value $\kappa_0$. The curvatures $\kappa_a$ and $\kappa_x$ are related as
\begin{equation}
\kappa_a \sin \frac{L(1+\epsilon/2)}{2 R_a}= R_x(1-\cos \frac{L}{2R_x})+\frac{L}{2}.
 \label{enflat1}
\end{equation}
This equation is transcendental, and therefore the optimum has to be computed numerically. We found that $1/R_x$ decreases with $1/R_a$ at any $\epsilon>\epsilon_c$; thus, equilibrium configuration is attained when driven filaments are at the optimal curvature $1/R_0$.

Rotating the driven filaments off-plane around their midpoint in opposite directions by some angle $\xi$ reduces the projection of the base of the driven filament on the original plane to $\widehat{d}_a=d_a\cos \xi$; as a result $h_x=\widehat{d}_a-d_y$ and curvature $1/R_x$ decreases (Fig.~\ref{AntiSymRaRxEps}a). However, $d_c$ also contracts with increasing angle $\xi$ as $d_c=\widehat{d}_x=\sqrt{d_x^2-(d_a \sin \xi)^2}$, which leads to buckling of the central perpendicular filament and a rise of bending energy. Off-plane rotation around the midpoint, in addition, causes twist in the central perpendicular filament. Due to the ensuing increase of energy, the planar state always remains optimal, as shown in Fig.~\ref{AntiSymRaRxEps}a. 

When $\epsilon>\epsilon_c$ and $\delta<0$, the equilibrium state is also planar (Fig.~\ref{frame}a) and, consequently, no twist arises. The parallel passive filament may bend in-plane or the driven filaments may reduce their curvature to lower the difference $d_y-d_a$. In one limit, the parallel passive filament remains straight, and the perpendicular filaments bend with the same curvature radius $R_x$ as at $\delta >0$, with the only difference that the framing filaments bend now outward. The dependence of optimal values of $R_a$ and $R_x$ on the extension coefficient $\epsilon$ is shown in Fig.~\ref{AntiSymRaRxEps}b. The optimal curvature of driven filaments grows with $\epsilon$, but bending resistance of passive filaments leads to an equilibrium at $1/R_a < 1/R_0$. The curvature $1/R_x$ attains a minimum at the critical elongation $\epsilon_c$ and then rapidly increases.  

\begin{figure}[b]
\begin{tabular}{cc}
		(a)&(b)\\
 \includegraphics[width=.23\textwidth]{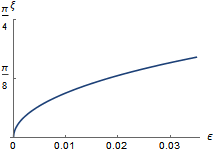}
 &\includegraphics[width=.23\textwidth]{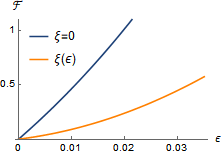}
 \end{tabular}
\caption{(a) The rotation angle of driven filaments around their midpoint as a function of the extension coefficient and (b) the energy gain of a twisted shape over the planar configuration in a frame with a symmetric arrangement of driven filaments }  
\label{frameS} 
\end{figure}

\subsection{Symmetric actuation}\label{S53}

When the curvature radii of driven filaments are oriented inside the frame, both central filaments remain straight (Fig.~\ref{frame}b,c). A configuration with $1/R_x\neq0$, $R_a = R_0$, satisfying $d_x=d_c-h_a$, with driven filaments remaining parallel, becomes not optimal and the frame switches to a twisted shape to reduce $d_x$. The twist angle $\xi$ can be found together with $h_x$ by solving the relations 
  \begin{equation}
h_x=|{d}_a\cos \xi-d_y|,  \qquad  d_x^2=(d_c-h_a)^2+(d_a \sin \xi)^2.
 \label{hxdx}
\end{equation}
The result depends on $R_a$ only if the central filaments are straight. The twist energy differs from zero in both configurations. In the first case, the torque is applied only to driven filaments, and at $\xi\neq0$ all filaments experience a twist.  We found that lowest energy configurations is attained at some optimal angle $\xi$ that minimizes the overall energy with $R_a>R_0$, and the twist angle depends on the extension $\epsilon$ (Fig.~\ref {frameS}a). The overall energies for shapes with $\xi=0$ and for twisted configurations at optimal $\xi(\epsilon)$ are shown in  Fig.~\ref {frameS}b. A twisted shape is preferable at any $\epsilon$, since twist provides a significant advantage by reducing bending energy. Typical shapes for configurations with parallel driven filaments and filaments turned by $\xi$ are presented, respectively, in Fig.~\ref{frame}b,c.

\section{Shape multistability in textiles}\label{S6}

\subsection{Reshaping of a rectangular strip}\label{S61}

\begin{figure}[t]
\begin{tabular}{cc}
		(a)&(b)\\
 \includegraphics[width=.26\textwidth]{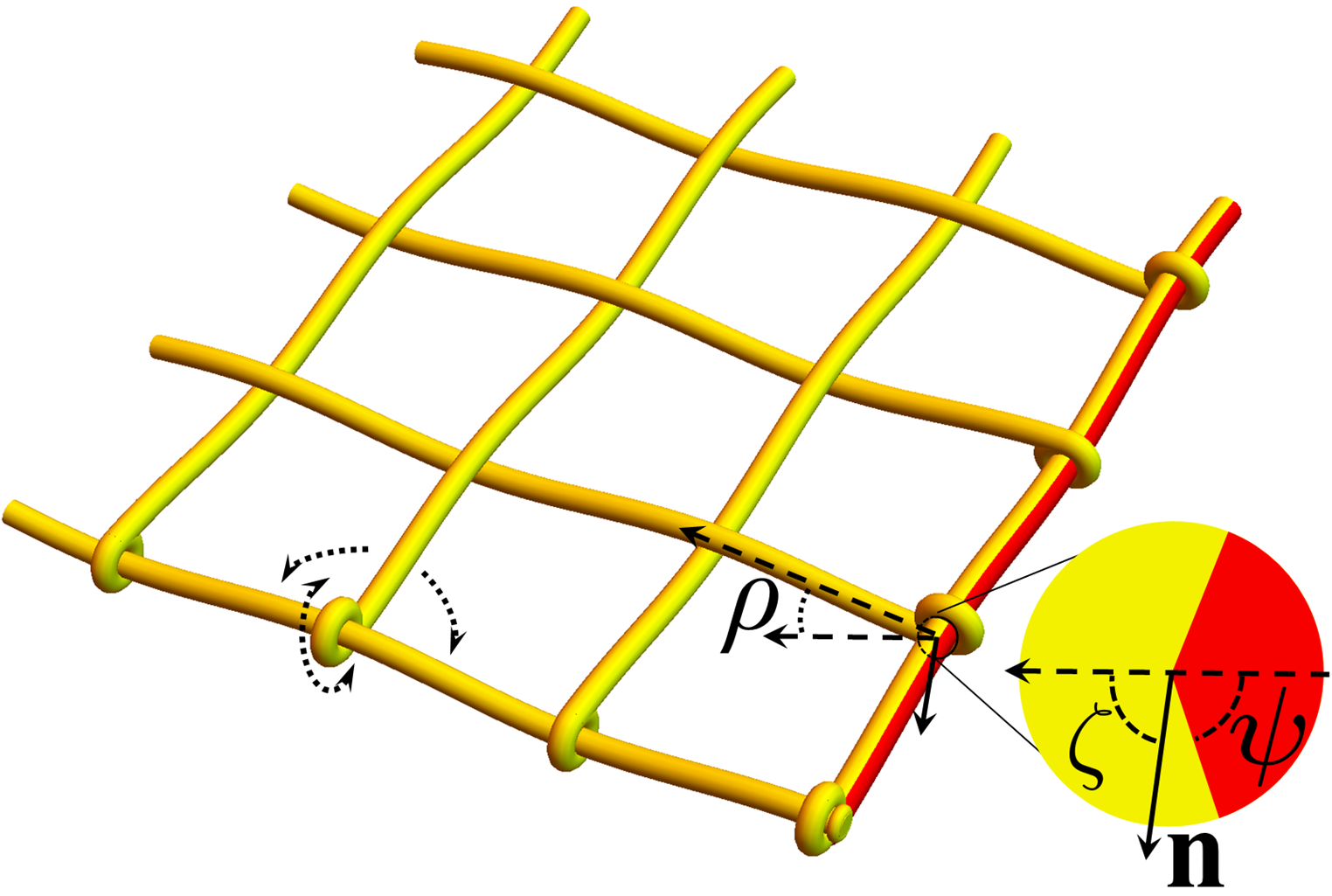}
 &\includegraphics[width=.22\textwidth]{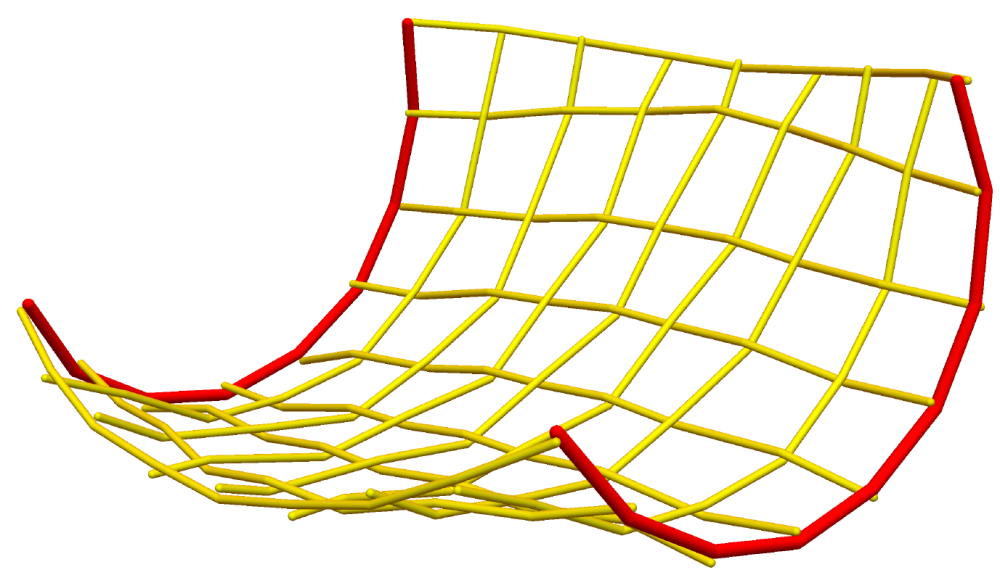}
\end{tabular}
\caption{(a) A piece of woven structure with loop connections at the ends and a Janus filament oriented at an angle $\rho$ to the midsurface. The cross-section of the Janus filament shows the driven sector with the angle $\psi$ and the deviation of normal vector \textbf{n} in the Frenet frame from the material normal by an angle $\theta$. (b) An equilibrium shape of textile with two untwisted Janus filaments that are allowed to freely rotate retaining the zero mismatch angle $\zeta$. }  
\label{ActFibRotation} 
\end{figure}

Full-fledged textiles present further examples of multistability. We start with reshaping of a planar rectangular piece of textile embroidered over opposite sides by two driven filaments. Interlaced passive filaments form the rest of fabric. All filaments have the same mechanical properties, but we assume that Janus filaments are twice as thick as the passive ones to make the reshaping effect more pronounced. To keep the integrity of the fabric, the structure should be stitched at the edges. Unlike the arrangement in our earlier publication \cite{textile}, we assume that internal filaments are stitched to the boundary filaments in a way allowing them to rotate around the latter's axis (see Fig.~\ref{ActFibRotation}a). The boundary filaments apply a torque, and cause a twist in stitched filaments when rotating off-plane. Since torque is applied at the ends only, twist is uniformly distributed along internal filaments and is determined by the angle difference between projections of border filaments onto the normal plane. In order to induce twist also in Janus filaments, we assume that they are attached in the same way to the framing passive fibers. This brings about a larger diversity of shapes compared to the case when driven filaments are free to rotate, leading to a bent equilibrium shape with no twist, as in Fig.~\ref {ActFibRotation}b.

Equilibrium configurations can be attained following pseudo-time evolution equations for the locations of the centerlines of each filament $\mathbf{x}_i(s)$ together with additional angular variables affecting twist. The positional equations are discretized as locations of the intersection nodes: 

 \begin{align}
\frac{d \mathbf{x}_{ij}}{d t}&= - \frac{\partial}{\partial \mathbf{x}_{ij}} 
 \sum_\textrm{segm}\mathcal{F}_\textrm{segm},
 \label{evold}
\end{align}
where the $\mathcal{F}_\textrm{segm}$ is the energy of filament segments between the nodes.
  
Altogether, there are three degrees of freedom at each node, i.e $3mn$ degrees of freedom in a rectangular piece of textile woven of $m+n$ filaments. In addition, there are $m+n-2$ variables describing the difference between orientations of the framing fibers at the two ends of all filaments (except the framing passive ones) that determine the respective torques. We solve the optimization problem for the elastic energy using the gradient descent method, starting from a reference state and applying small local perturbations by each degree of freedom to compute energy gradients. After all gradients are found, the node positions and angles are simultaneously updated and the procedure is repeated.     

The woven structure of the textile, depicted in Fig.~\ref{ActFibRotation}a, imposes a ``microcurvature" component along the normal to the envelope surface of the fabric, that depends on the distance $\ell_{i,j}$ between intersections and prevents convergence of neighbouring nodes to distances comparable to the diameter of the filaments. The microcurvature $\kappa_{i,j}$ at a node $(i,j)$ of the $i$th filament is approximated in the limit $r \ll \ell_{i,j}$ by $\kappa_{i,j} \approx 4r/(\ell_{i,j+1}+\ell_{i,j-1})$.  In actual computations, this additional curvature is taken into account by shifting the locations $x_{i,j}$ of intersecting filaments by $r$ along the normal to the envelope of the fabric and using the approximant $\kappa_{i,j} = |\alpha_{i,j}| /\min<\ell_{i,j+1},\ell_{i,j-1}>$, where $\alpha_{i,j}$ is the angle between the two adjacent links. The approximation is valid as long as the local curvature radius is large compared to the distance between nodes. The intersecting filaments should remain in contact under these conditions, since their separation will increase the microcurvature, which is minimal (though non-zero) when the network is regular and increases with growing inhomogeneities.

 \subsection{Antisymmetric actuation}\label{S62}

  \begin{figure}[b]
  	(a)	\\
  \includegraphics[width=.3\textwidth]{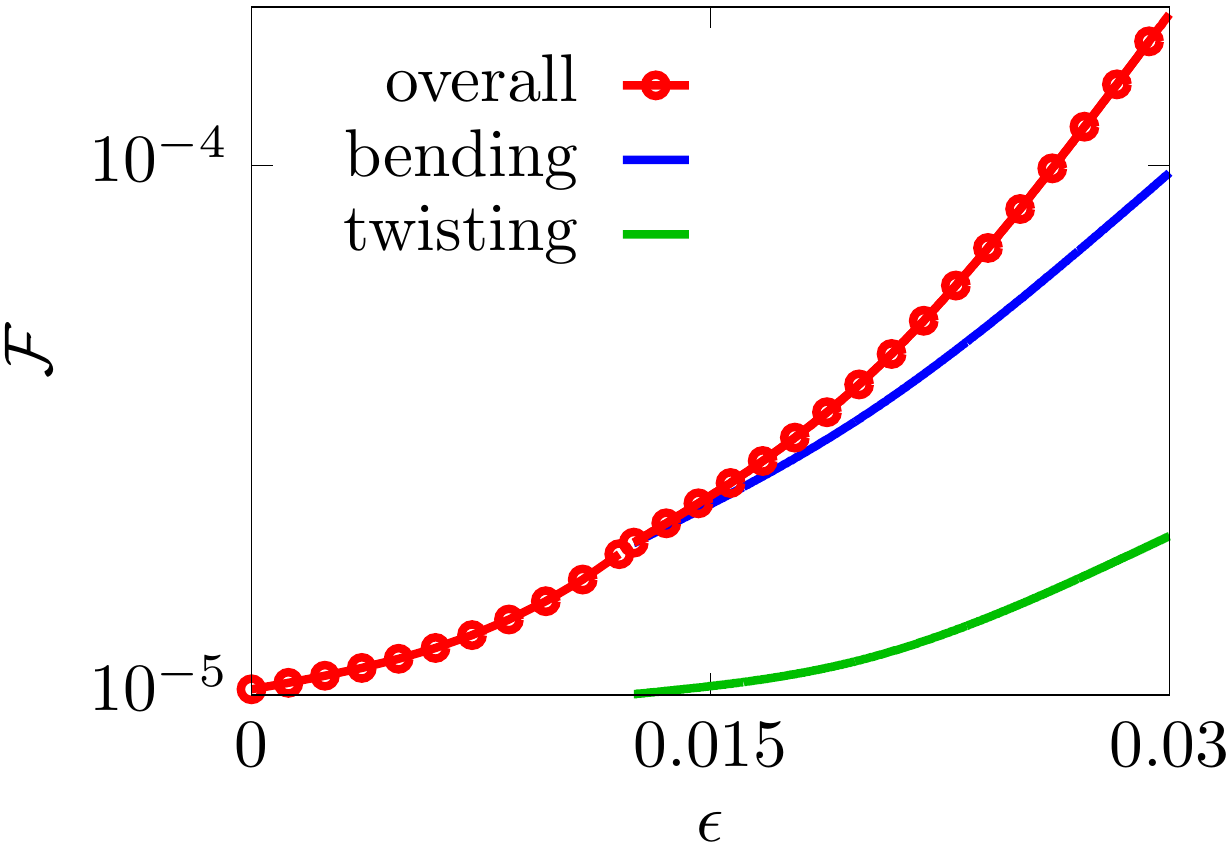}
  \begin{tabular}{cc}
  		(b)&(c)\\
   \includegraphics[width=.23\textwidth]{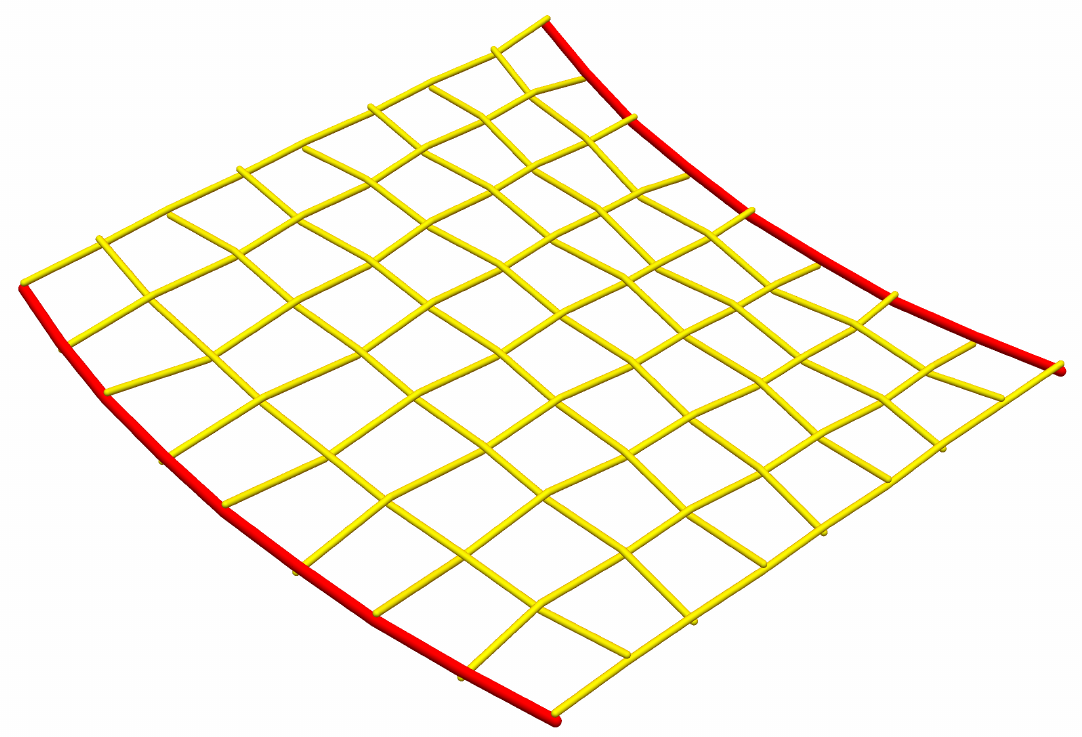}
  &\includegraphics[width=.24\textwidth]{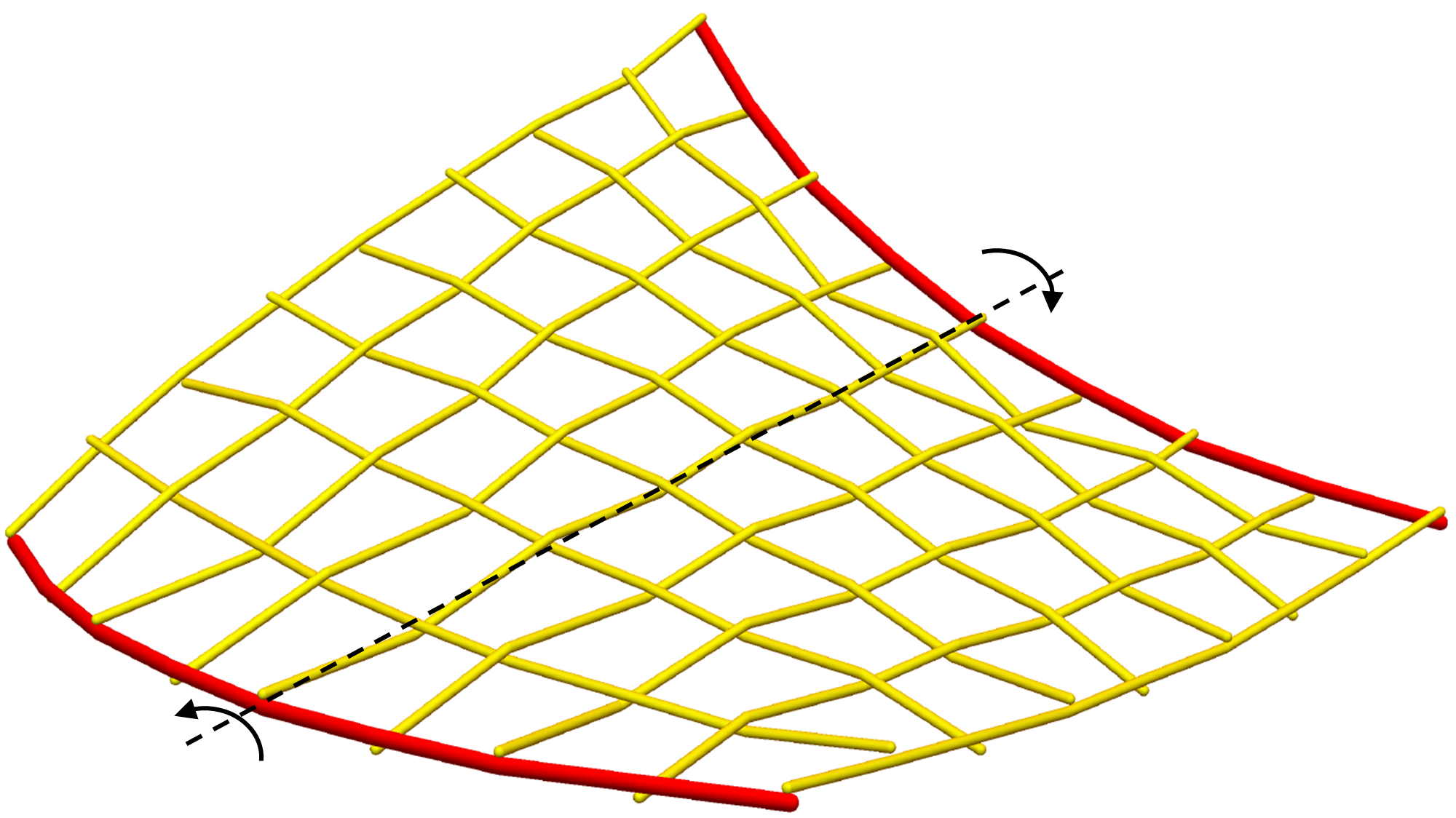}
  \end{tabular}
  \caption{Antisymmetric arrangement of textile. (a) Free energy as a function of extension coefficient $\epsilon$. (b) A flat textile structure at $\epsilon=0.0125$. (c) A twisted textile structure at $\epsilon=0.0175$}  
  \label{Textile_0_0} 
  \end{figure}

An initially planar textile with two framing Janus filaments, which are oriented in the way causing them to bend in the same direction in the textile plane, remains flat until $\epsilon\approx0.0125$ (Fig.~\ref{Textile_0_0}). No twist arises in such a configuration; at $\epsilon<\epsilon_c$ the Janus filaments elongate rather than develop curvature, leading to a slight bend of boundary passive filaments. At $\epsilon_c<\epsilon<0.0125$, the perpendicular passive filaments remain straight, changing angle at connections with Janus filaments, but parallel passive filaments bend in the enveloping plane of the textile because the distance between the ends of the Janus filaments shortens. Since one of driven driven filaments bends inward, becoming closer to parallel filaments, and another one bends outward, increasing distance from the neighboring filament, bending is not symmetric and depends on the filament position. The parallel filaments in the center of textile remain almost straight due to a lower compression, while filaments near the driven ones bend in the same direction to reduce microcurvature and to prevent convergence of filaments. As a result, the structure has different density of parallel passive filaments, as seen in Fig.~\ref{Textile_0_0}b.           
  
As the actuation rate increases, the difference in density, as well as in rigidity, across the textile grows, leading to a difference between the curvatures of driven filaments. It becomes more difficult to attain the intrinsic curvature for the Janus filament that bends inward. The distance between the ends of the driven fibers becomes shorter than between their middle points, and therefore perpendicular passive filaments are compressed near the boundaries. This incompatibility grows with $\epsilon$ and, finally, the structure bends out of plane and reshapes to a twisted configuration of two alternative orientations (Fig.~\ref{Textile_0_0}c). This causes an increase of distances between the ends of the driven filaments which moderates bending of passive filaments. This shape is preferable because twisting energy is less costly than bending energy. Moreover, such an arrangement is only slightly affected by twisting rigidity, and equilibrium structures at different $\epsilon$ qualitatively repeat our previous results in Ref.~\cite{textile}, where filaments were allowed to rotate about their axis at the boundary nodes. Since in an elementary frame with the same arrangement of actuation the difference in curvatures of driven filaments does not arise, the frame, unlike textile, remains planar, and there is no reason to switch it to a twisted configuration.

    \begin{figure}[t]
    (a)	\\
    \includegraphics[width=.3\textwidth]{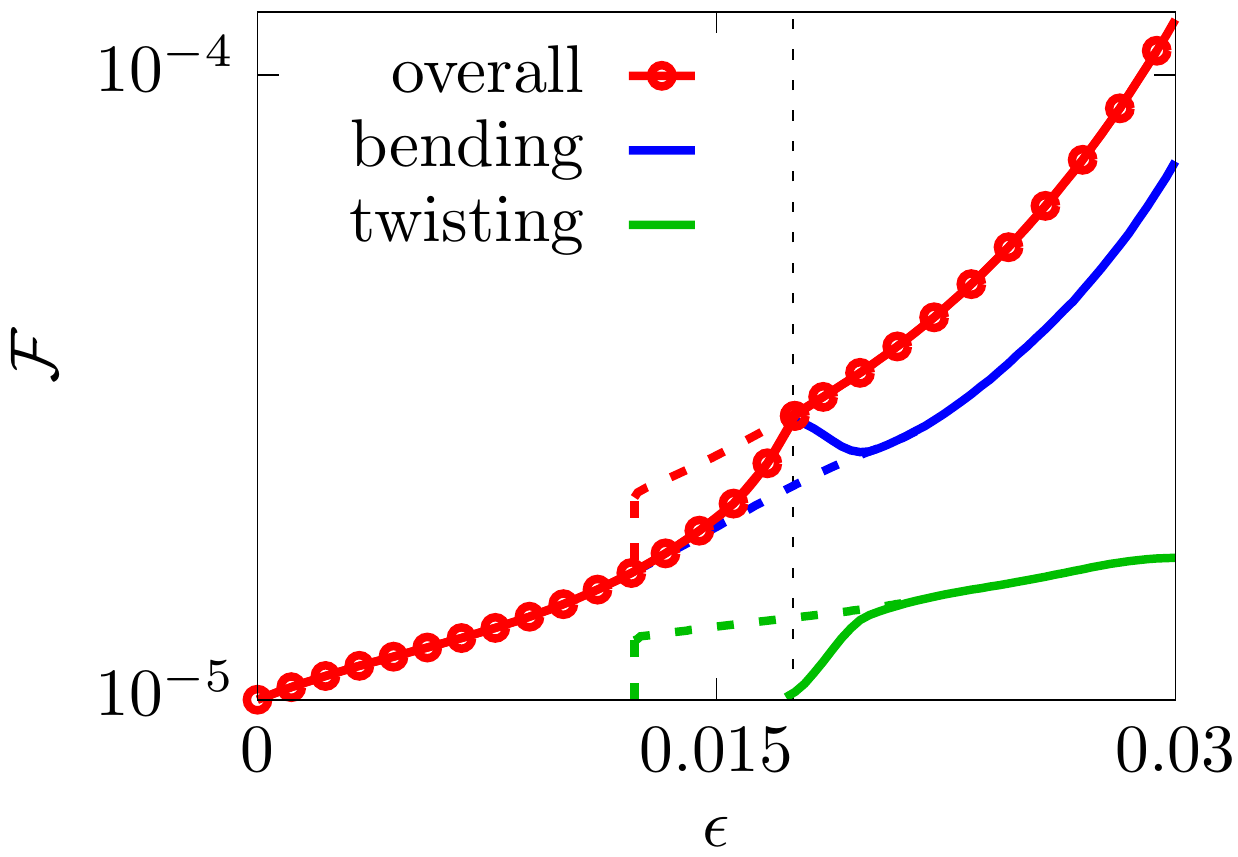}
    \begin{tabular}{cc}
    		(b)&(c)\\
     \includegraphics[width=.24\textwidth]{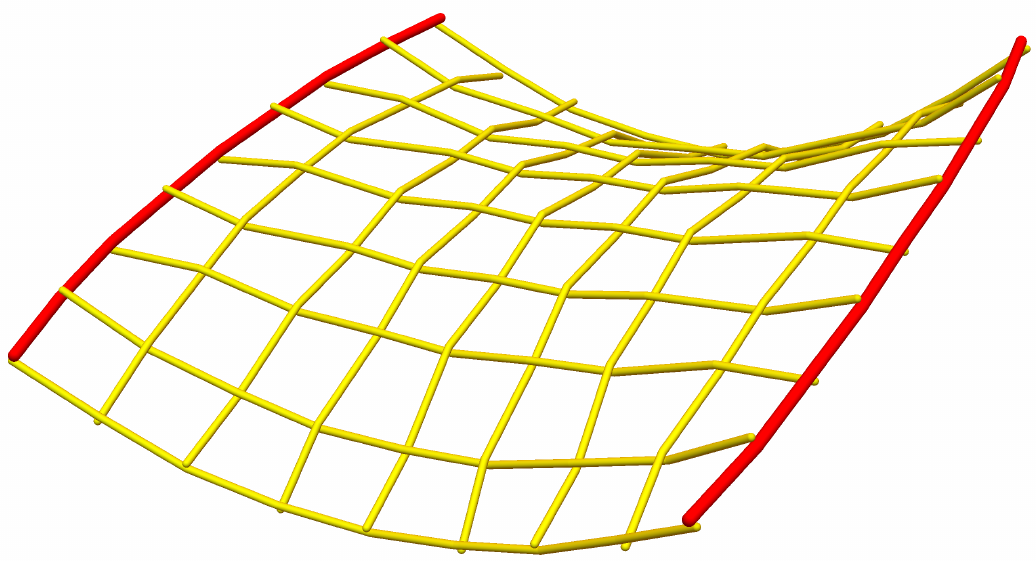}
    & \includegraphics[width=.18\textwidth]{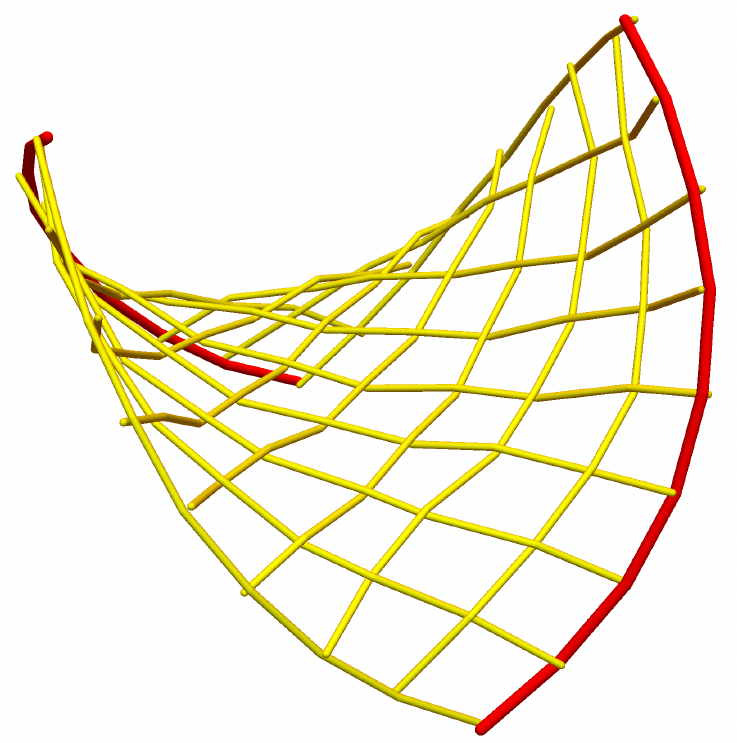}
    \end{tabular}
    \caption{Symmetric actuation of a textile piece. (a) Free energy as a function of the extension coefficient $\epsilon$. Stable and metastable states are plotted by solid and dashed lines respectively. The vertical dotted line shows the critical value of $\epsilon$ at the shape transition. (b,c) Typical configurations, respectively, before ($\epsilon=0.015$) and (c) after ($\epsilon=0.02$) the shape transition.}  
    \label{Textile_0_pi} 
    \end{figure}
  \subsection{Symmetric actuation}\label{S63}
 
The trade-off between bending and twisting energies becomes more pronounced in a symmetric arrangement where no planar configurations exist at $\epsilon \neq 0$. When the extension coefficient is small, the Janus filaments develop curvature by bending the entire structure, which increases distortion close to the passive boundaries for perpendicular filaments and close to the middle line, for parallel ones (Fig.~\ref{Textile_0_pi}b). The Janus filaments develop intrinsic curvature, and yet they retain the symmetry with no twisting energy arising in such configuration up to $\epsilon < 0.0175$.  However, the resistance to bending of the passive part of textile leads to a fast growth of the bending energy, which is reduced by transition to a twisted shape (Fig.~\ref{Textile_0_pi}c), since it is less energetically expensive to twist than to bend a rod. The structure evolves by rotating about the middle line, increasing the distances between the ends of the driven fibers. As a result, it reduces compression of passive fibers and the bending energy, but induces twisting energy costs. Such a shape change occurs at $\epsilon\approx 0.0175$, accompanied by a jump of twisting and bending energies clearly distinguished in the energy plot (Fig.~\ref{Textile_0_pi}a).

The reverse transition to a small $\epsilon$ and non-twisted configurations, takes place at lower value $\epsilon\approx0.012$. It forms a hysteresis loop with a coexistence interval of the untwisted configuration, which is absolutely stable and has a lower overall energy, and the twisted one, which is metastable and exists within the interval $0.012<\epsilon<0.0175$ and can be attained only from a twisted shape at higher $\epsilon$. We note that the textile shapes at $\epsilon>0.0175$ reproduce the optimal configurations found for a frame of the same type of actuation.

\subsection{Imbedded Janus rings}\label{S71}
 
Janus filaments forming closed loops provide neat examples of elastic instabilities  by generating a variety of elaborate shapes. Isolated twisted closed filaments are absolutely unstable to off-plane deformations and acquire multiple convoluted shapes \cite{pre18}, but they can be stabilized when restricted by passive filaments or woven into a textile fabric. Such arrangements provide ways to control the reshaping and to extend understanding of connectivity in woven structures.    
 
We consider first the shape change of a twisted Janus ring connected by two passive filaments crossing in the center with the lengths equal to the ring diameter, which is initially in a planar configuration. The passive filaments are not allowed to change positions of their connecting points with the Janus filament, but they can rotate about it and change the angle at these nodes. Rotation of the Janus filament around the centerline of passive filaments imposes, however, a twist, and the reshaping leads to compression of passive fibers. We assume that the central node may change its position, but the filaments remain in contact, as it was also assumed for an elementary frame in Sect.~\ref{S5}.
 
 \begin{figure}[b]
  		\begin{tabular}{cc} 
  			(a)&(b)\\
  			\includegraphics[width=.235\textwidth]{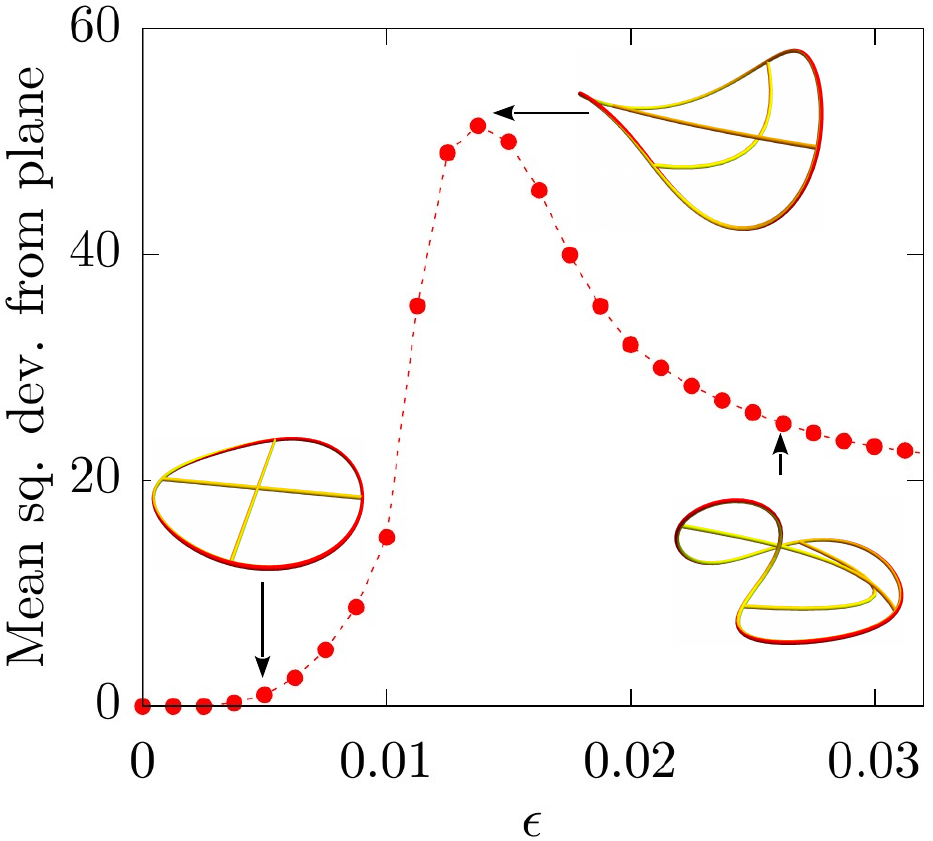}
  			&\includegraphics[width=.235\textwidth]{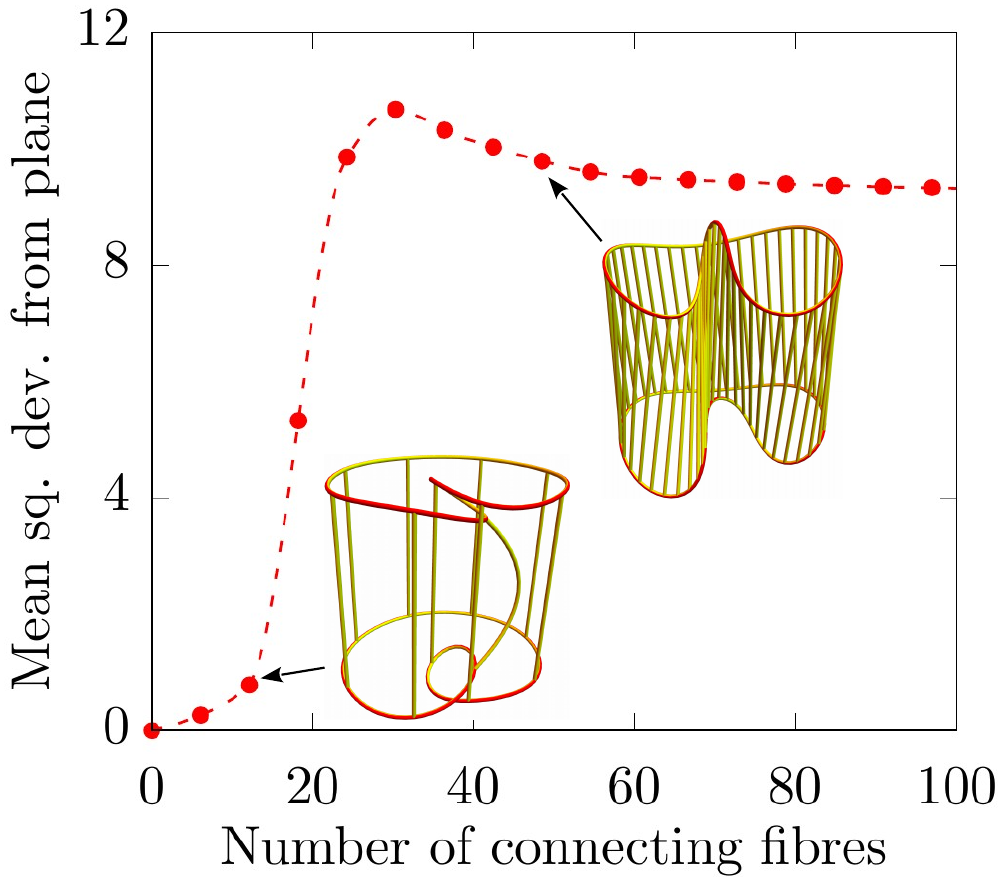}
  		\end{tabular}
  		\caption{(a) Reshaping of a twisted Janus ring restricted by two perpendicular passive filaments. (b) Deformation of two twisted Janus rings with opposite coiling handedness connected by a different number of passive filaments. }
  		\label{Restriction}
  	\end{figure} 
  
A reshaping ring passes through at least three qualitatively different configurations as the extension coefficient increases. Deviation of the Janus filament from the original plane depends on the parameter $\epsilon$ and shown in Fig.~\ref{Restriction}a. At $\epsilon<0.005$, the bending energy of the ring is low compared to the twisting and bending energy of passive filaments, and the structure remains planar with minor deviations of the ring from initial circular shape. A further increase of $\epsilon$ leads to bending of the passive filaments. First, at $0.005<\epsilon<0.02$, only one of them buckles, while the other one remains straight, causing the ring to acquire a significantly non-planar shape. With further increase, the entire structure inflects, so that at $\epsilon>0.02$ there are no stable configurations with  passive filaments remaining straight. Finally, the Janus ring produces an additional loop, as it develops an internal twist to reduce the bending energy. The average deviation from original plane decreases but, unlike an unconstrained Janus ring \cite{pre18}, it cannot fold up to a planar configuration, and warps until being restricted by a self-intersection limiting further deformation.

Interaction between twisted closed Janus filaments connected by passive filaments also restricts their reshaping. This affects  shape transformations in a structure consisting two Janus rings of the same circumference and pitch length, but opposite coiling handedness. Fig.~\ref{Restriction}b compares structures with the same value of the extension coefficient $\epsilon=0.03$ but different number of connecting passive filaments. A single ring with an imposed twist by $2\pi$ folds at this value of $\epsilon$ to a double-covered circle, but connecting passive filaments constrain the deformations. When the total bending energy of connecting filaments is lower than the energy of Janus rings, each ring changes its shape to a form of almost flat partially folded loop. Increasing the number of passive filaments enforces the connectivity that prevents bending of passive filaments and leads to growing deviations of the rings from the planar configuration, forming  loops that are perpendicular to the original plane. Typical shapes at different values of connectivity are shown in Fig.~\ref{Restriction}b together with the mean squared deviation of the rings from the average vertical position.

\begin{figure}[t]
 		\begin{tabular}{ccc} 
 			(a)&(b)&(c)\\
 			\includegraphics[width=.155\textwidth]{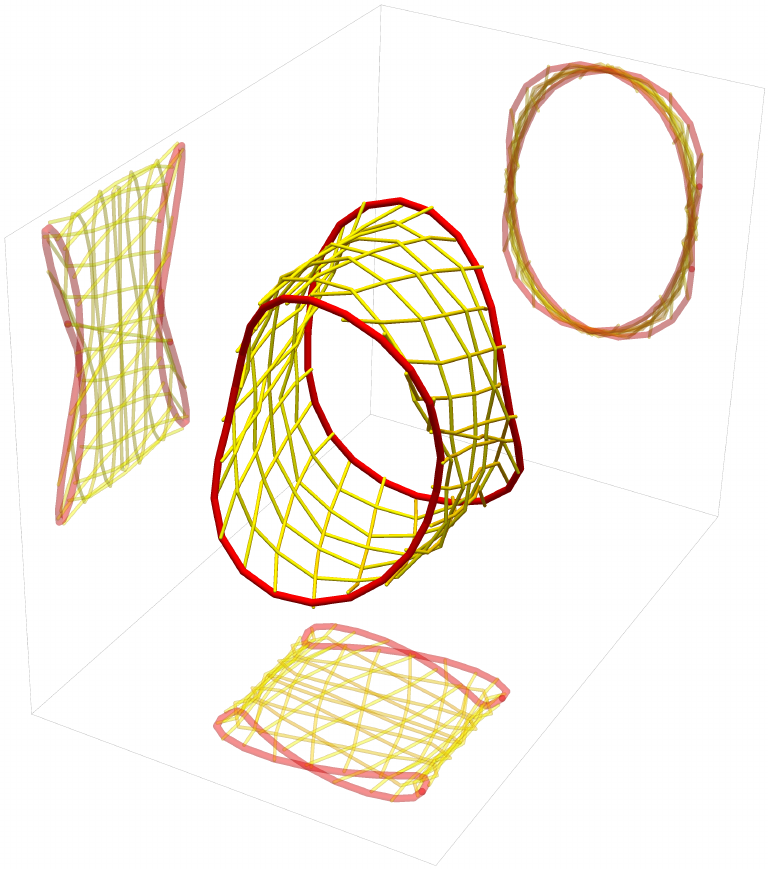}
 			&\includegraphics[width=.155\textwidth]{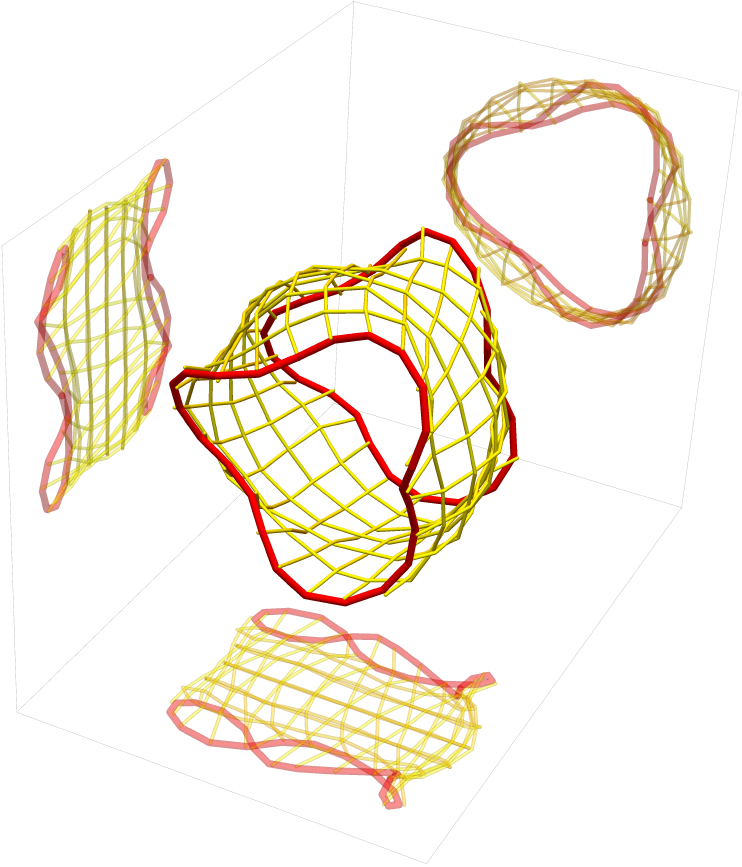}
 			&\includegraphics[width=.155\textwidth]{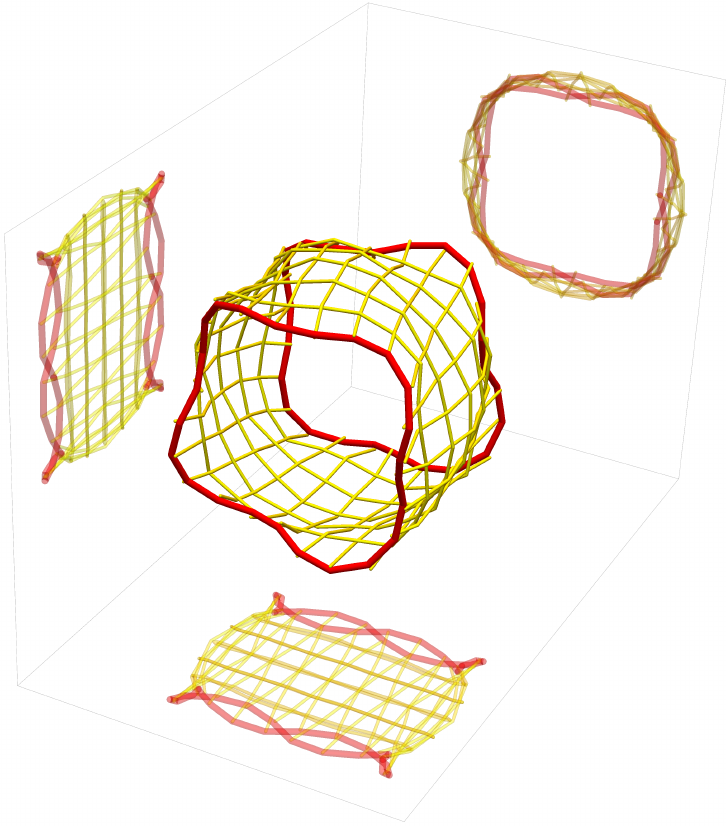}
 		\end{tabular}
 		\caption{Shapes of a cylindrical textile with two Janus filaments of the same handedness and (a-c) 2, 3, 4 turns of the basis, respectively, at the same extension parameter $\epsilon=0.015$. Projections on the three normal planes are also shown.}
 		\label{CylinderP}
 	\end{figure}

The change of reshaping regimes of constrained Janus rings becomes still more pronounced in woven cylindrical ``sleeve" structures. The results of computations for two identical Janus rings of the same handedness and pitch embroidered into a cylindrical textile are presented in Fig.~\ref{CylinderP}. The connecting passive filaments tend to damp the deformations, and therefore the two Janus filaments, unlike unconstrained twisted Janus rings, are prevented from attaining an optimal curvature by producing multiple loops. This causes a cylindrical fabric to deform into an ellipsoid with triangular,  square, etc. cross-sectional shapes, depending on the number of full internal rotations of the Janus ring. 
    
 \section{Non-uniform actuation}\label{S7}
\begin{figure}[p]
			\begin{tabular}{c} 
			(a)\\
			\includegraphics[width=.475\textwidth]{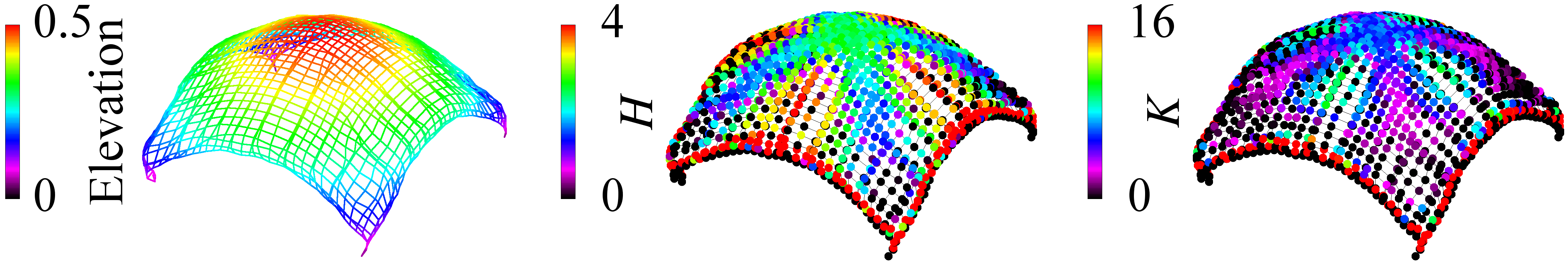}\\
			(b)\\
			\includegraphics[width=.475\textwidth]{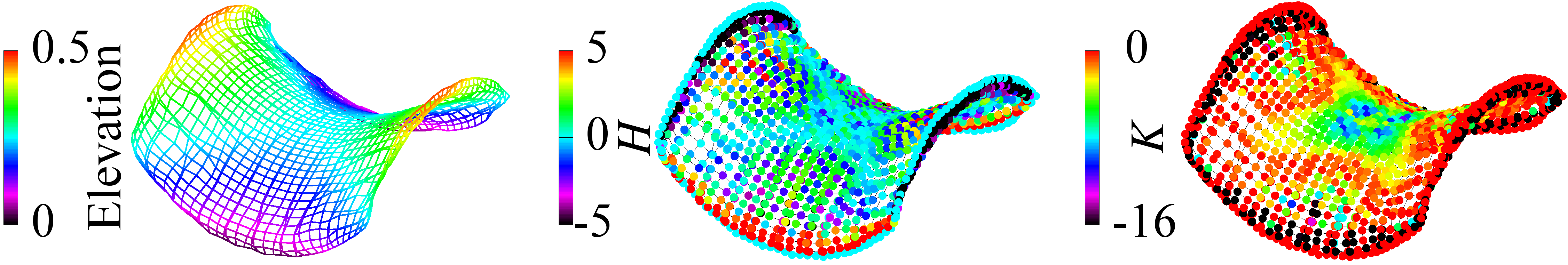}\\
				(c)\\
		\includegraphics[width=.45\textwidth]{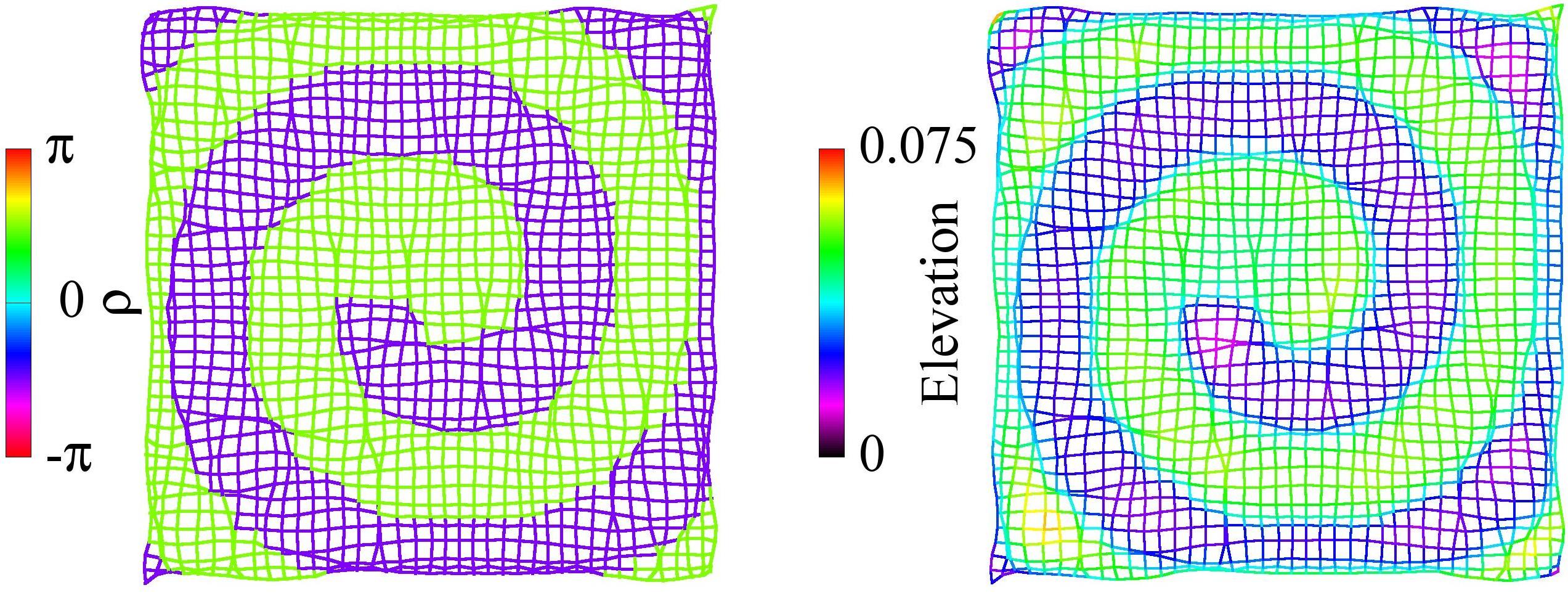}\\
			(d)\\
			\includegraphics[width=.45\textwidth]{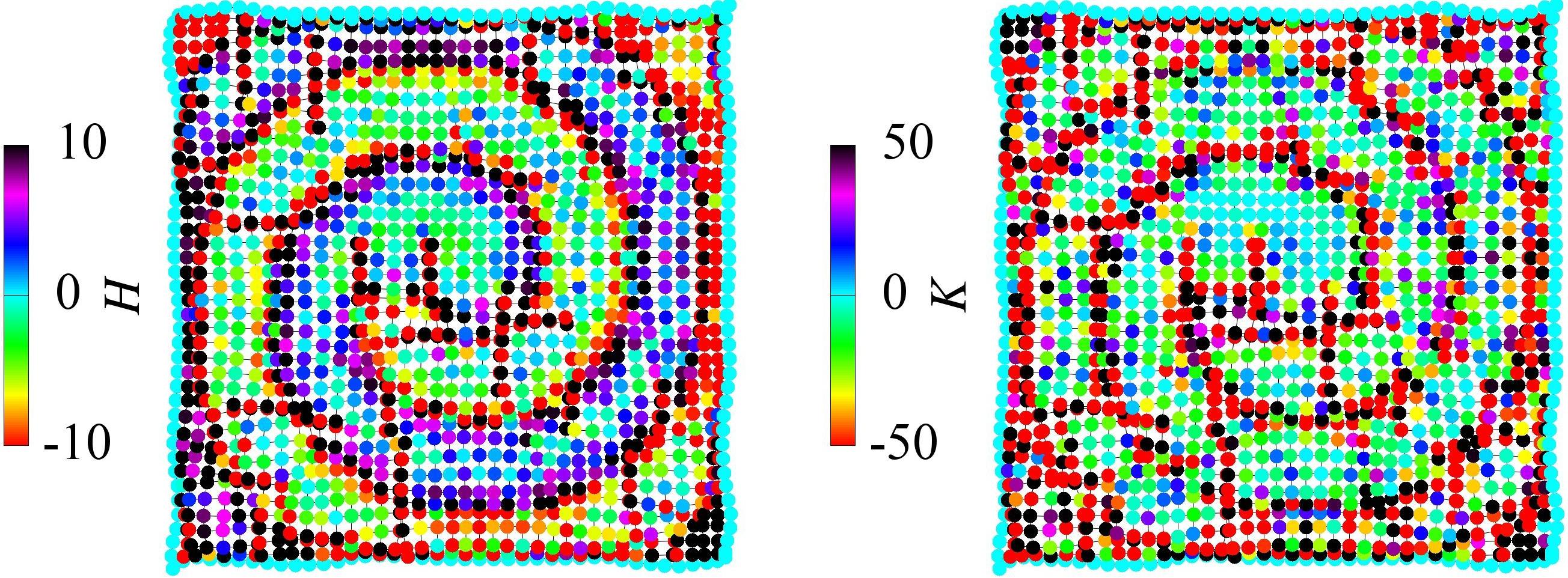}\\
			(e)\\
			\includegraphics[width=.45\textwidth]{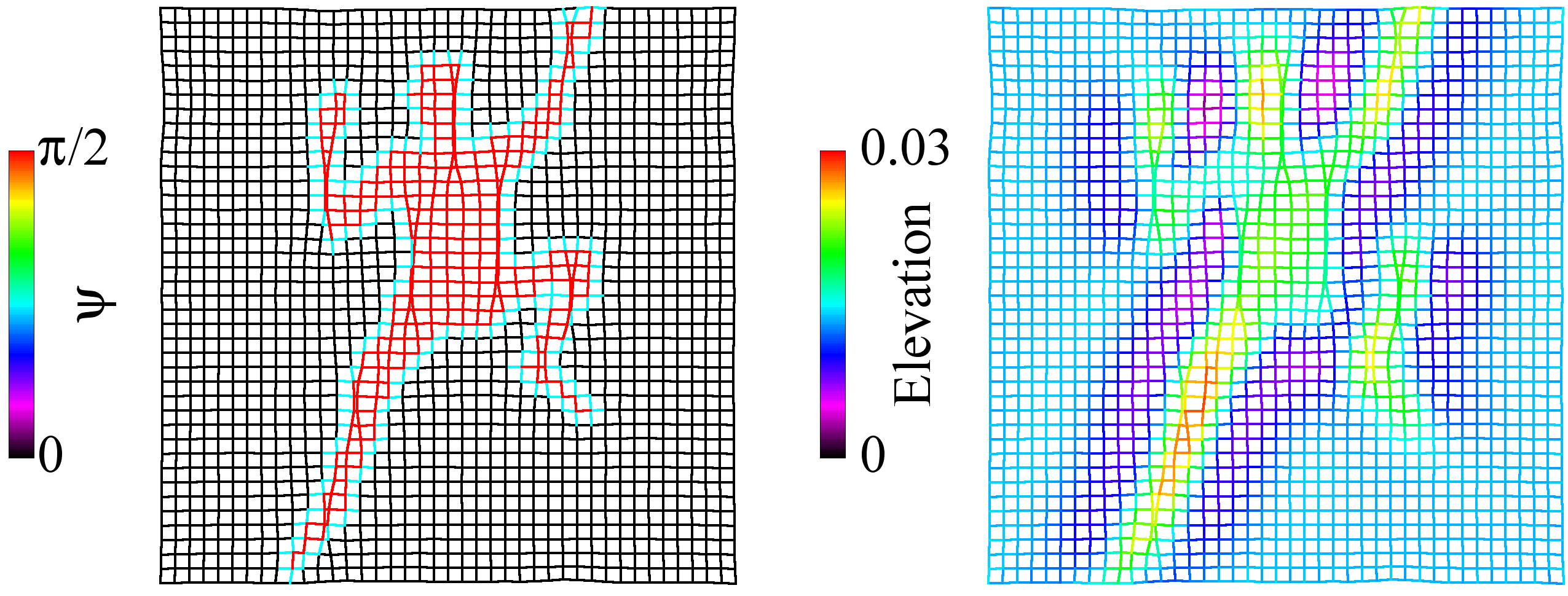}\\
			(f)\\
			\includegraphics[width=.45\textwidth]{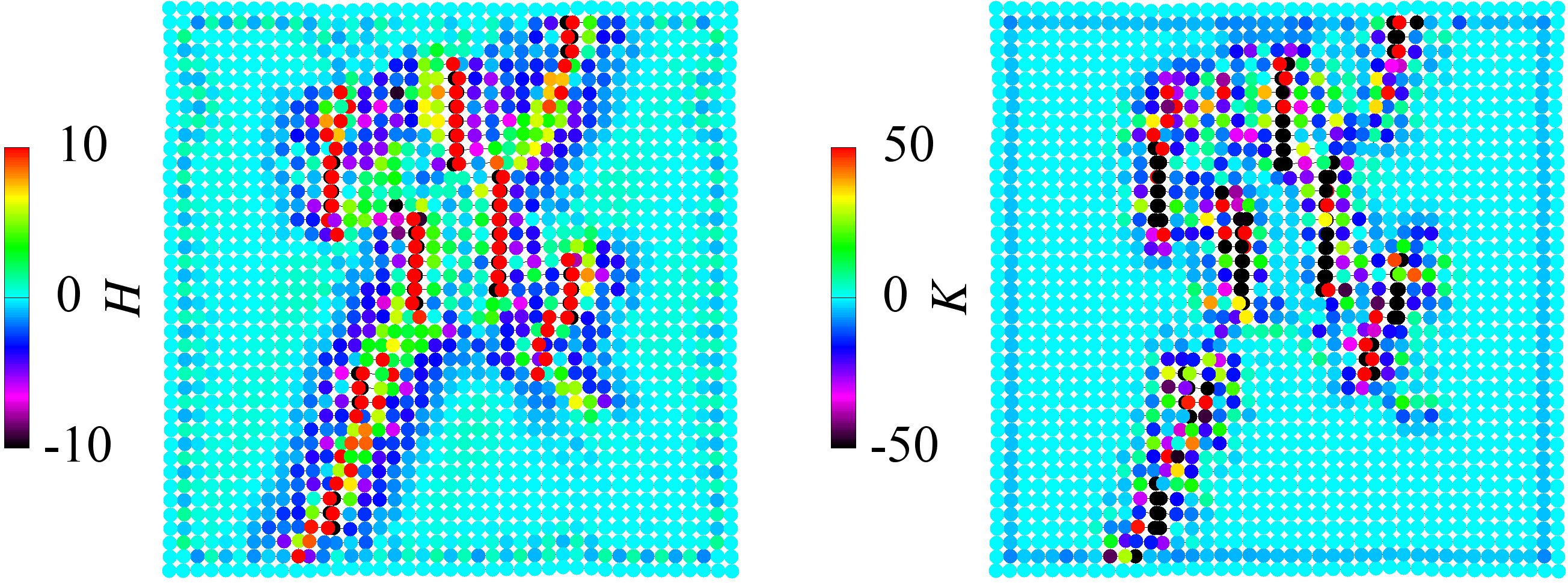}\\
			(g)\\
			\includegraphics[width=.475\textwidth]{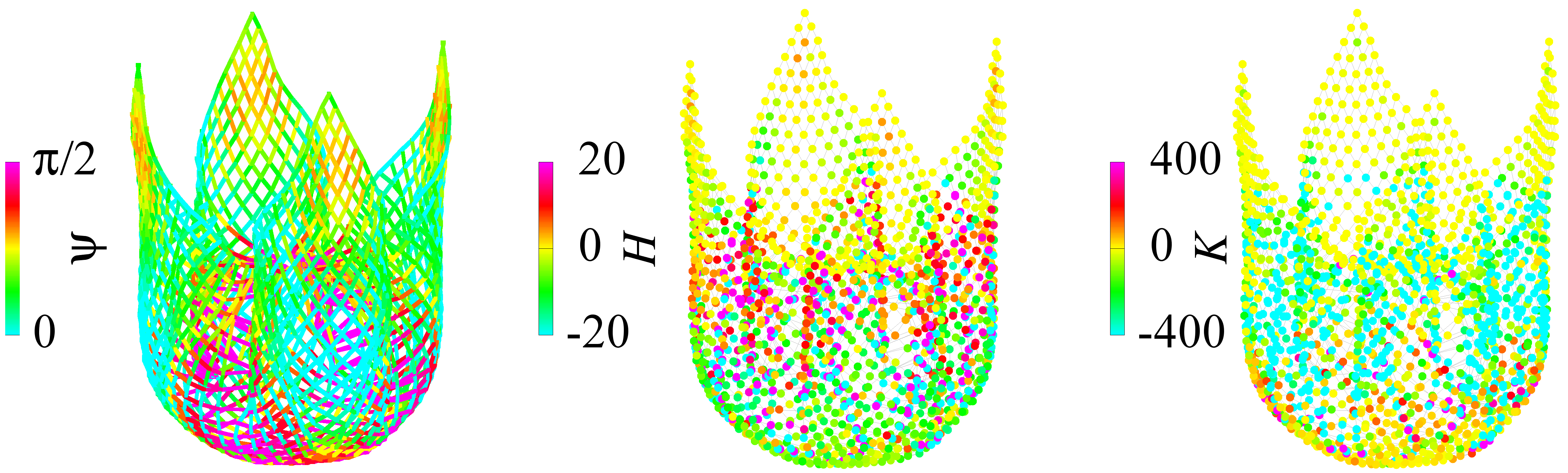}

						\end{tabular}
			\caption{Woven structures obtained by actuating a unit square piece of textile. (a,b): Shapes with positive and negative Gaussian curvature. (c,d): A spiral shape with constant areas of driven sectors. (e,f): A relief of a man at constant orientation of driven sectors. Colouring or shading shows the mean curvature $H$ (left) and the Gaussian curvature (right) at the nodes (a,b,d,f) or the orientation and elevation of filament sectors (c,e). (g) A flower shape obtained from textile with non-uniform internal orientation and amplitude of curvature. Parameters for (a-g): $\epsilon=0.03, \, r=10^{-3}$}
			\label{PosNeg}
\end{figure} 

We now extend our investigation to address the reshaping of woven structures with multiple Janus filaments of different intrinsic curvature and seek to understand the design criteria for creation of tunable textiles. Such metamaterials are sufficiently flexible to program on-demand internal geometry of each filament constituting the fabric and to to control desired morphing from 2D to 3D strictures.      
 
An initially planar piece of textile, unlike a continuous flat sheet, can develop shapes with a non-zero Gaussian curvature even when the driven sectors of each filament have a constant size and orientation along filaments. Examples of shapes with positive and negative Gaussian curvature are shown in Fig.~\ref{PosNeg}(a,b). In the first case, all filaments are oriented uniformly and tend to develop curvature normally to the textile surface, generating positive mean and Gaussian curvatures of the enveloping surface at each node. In the second case, negative Gaussian curvature is attainable when the filaments have opposite orientation in two orthogonal directions. The amplitudes of curvatures are, however,  not homogeneous at different nodes in both cases, reaching a maximum near the boundaries and being more uniform in the center of textile. Developing a non-zero Gaussian curvature causes compression and extension of different parts of structure, so that the initial regular arrangement becomes perturbed by convergence and dispersing of filaments upon transition to a new equilibrium shape.

A variety of 3D shapes can be obtained upon actuation when either areas or orientations of driven sectors, or a combination of both, are chosen in a slightly more artful way to reach a desired shape. For example, in Fig.~\ref{PosNeg}c,d, we define local orientation of the intrinsic curvature for each filament in such a way that generates a spatial spiral relief upon actuation. All filaments tend to develop curvature normally to the enveloping surface, but the structure has two domains of opposite orientation of intrinsic curvature, upward ($\rho=\pi/2$) and downward ($\rho=-\pi/2$) through reversing positions of the driven sectors in transitional zones, while the amplitude of internal curvature defined by the values of $\psi$ and $\epsilon$ is set constant in whole structure. As a consequence, the textile generates appropriate vertical out-of-plane displacements and non-zero Gaussian curvature of the enveloping surface.

Alternatively, the target shape can be attained by patterning the areas of driven sectors (i.e the angle $\psi$), while keeping the orientation of internal curvatures uniform along filaments. This generates local bending and normal displacement in the driven domain, while other nodes remain planar . In Fig.~\ref{PosNeg}e,f we demonstrate such a kind of actuation with a non-symmetric pattern of passive/driven sectors revealing a sculpture of climbing man. The resolution and details of the textures depend on the textile density and geometry, since the local curvature radius is assumed to be larger than the distance between filaments.

A combination of non-uniform driven sectors and variable orientation of intrinsic curvatures provide a flexibility in design of more complex 3D shapes. In order to generate a shape that perfectly recovers a target 3D object, we must solve a non-trivial inverse design problem. The solutions provided here are imperfect, but our aim is to demonstrate advantages of shape-programmable textile capable to morph in a sophisticated way to reproduce desired shapes, even without solving the inverse problem precisely. In particular, we prescribe an arrangement that generates a flower shape, which is shown in Fig.~\ref{PosNeg}g. This complex shape cannot be attained by prescribing either $\psi$ or $\rho$, since both the amplitude and direction of curvature vary at each node. The orientation of filaments throughout the structure are chosen to change gradually from $\rho=0$ at one boundary to $\rho=\pi$ at opposite edge, so that the boundary filaments develop curvature directed inward and normally to enveloping surface in the center. The amplitude of the curvature is defined in such way that renders "flower petals" locally flat and oriented upward.

\section{Conclusion}

Morphing driven Janus filaments and textiles comprising various combinations of driven and passive filaments creates a wide variety of shapes. Transitions between qualitatively different structures minimizing the overall bending and twisting energy may take place at certain values of the extension coefficient, and metastable structures may coexist with an absolutely stable state under certain conditions. Reverse design of desired forms can be achieved by programming the size and orientation of a driven sectors in Janus filaments; this task becomes both more flexible and more difficult, compared with either single filaments or continuous sheets when morphing textile composed of or embroidered with Janus filaments.

\emph{Acknowledgement} This research is supported by Israel Science Foundation (grant No. 669/14). 


\begin{thebibliography}{90}

\bibitem{robot}D. Rus and M. T. Tolley, Design, fabrication and control of soft robots, Nature \textbf{521} 477--475 (2015).
\bibitem{smart}E. X. Tao, Handbook of Smart Textiles (Springer, Singapore, 2015).
\bibitem{Balazs}V. V. Yashin, O. Kuksenok, P. Dayal, and A. C. Balazs, Rep. Progr. Phys. \textbf{75} 066601 (2012).
\bibitem{Warner} M. Warner and E. M. Terentjev, \emph{Liquid Crystal Elastomers} (Clarendon Press, Oxford, 2003).
\bibitem{sharon} H. Aharoni, E. Sharon, and R. Kupferman, Phys. Rev. Letts. \textbf{113}, 257801 (2014).
\bibitem{mostajeran} C. Mostajeran, M. Warner, T. H. Ware, and T.J . White, Proc. R. Soc. A \textbf{472}, 20160112 (2016).
\bibitem{Aharoni}H. Aharoni, Y. Xia, X. Zhang, R. D. Kamien, and S. Yang, PNAS \textbf{115} 7206 (2018).
\bibitem{Goriely06} A. Goriely, J. Elasticity \textbf{84}, 281 (2006).
\bibitem{GorielyTabor} A. Goriely and M. Tabor, Nonlinear Dyn. \textbf{21}, 101 (2000).
\bibitem{GorielyGoldstein} R. E. Goldstein and A. Goriely, Phys. Rev. E \textbf{74}, 010901 (2006).
\bibitem{Ionov}L. Ionov, G. Stoychev, D. Jehnichen, and J. U. Sommer, ACS Applied Materials \& Interfaces \textbf{9} 4873 (2017).
\bibitem{Goriely02}T. McMillen and A. Goriely, J. Nonlinear Sci. \textbf{12}, 241 (2002).
\bibitem{wiki}https://en.wikipedia.org/wiki/Wire{\textunderscore}sculpture.
\bibitem{Tobias}I. Tobias and W. K. Olson, Biopolymers, \textbf{33}, 639-646 (1993).
\bibitem{pre18} A. P. Zakharov and L. M. Pismen, Reshaping of a Janus ring, Phys. Rev. E \textbf{97}, 062705  (2018).
\bibitem{textile} A.~P. Zakharov and L.~M. Pismen, 
Soft Matter \textbf{14}, 676 (2018).
\bibitem{pre17} A. P. Zakharov and L. M. Pismen, Phys. Rev. E \textbf{96}, 012709  (2017).
\bibitem{Maha}A. S. Gladman, E. A. Matsumoto, R. G. Nuzzo, L. Mahadevan, and J. A. Lewis, Nature Mater. 15, 413 (2016).

\end{thebibliography}
\end{document}